\documentclass[a4paper,aps,prd,10pt,preprintnumbers,showpacs,twocolumn,superscriptaddress,nofootinbib,amsmath,amssymb,gensymb,amsfonts]{revtex4-1}
\usepackage{graphicx}
\usepackage{cmap}
\usepackage[utf8]{inputenc}
\usepackage[T1]{fontenc}
\usepackage{nicefrac}
\usepackage{bm,amsmath,amssymb,amsfonts,gensymb,bbold}
\usepackage[colorlinks=true,linktocpage=true,linkcolor=blue,citecolor=blue,allcolors=blue]{hyperref}
\usepackage{amsmath}
\usepackage{makecell}
\usepackage{soul}
\def\imo{i}
\def\re#1{Re(#1)}
\def\im#1{Im(#1)}
\def\K{{\cal K}}

\newcommand{\eq}[1]{\begin{align} #1 \end{align}}
\begin{document}
\title{Probing the Effective Quantum Gravity via Quasinormal Modes and Shadows of Black Holes}
\author{R. A. Konoplya}\email{roman.konoplya@gmail.com}
\affiliation{Research Centre for Theoretical Physics and Astrophysics, Institute of Physics, Silesian University in Opava, Bezručovo náměstí 13, CZ-74601 Opava, Czech Republic}
\author{O. S. Stashko}\email{alexander.stashko@gmail.com}
\affiliation{Princeton University, Princeton, NJ, 08544, USA}
\begin{abstract}
Two quantum-corrected black hole models have recently been proposed within the Hamiltonian constraints approach to quantum gravity, maintaining general covariance \cite{Zhang:2024khj}. We have studied the quasinormal spectra of these black holes using four methods: the higher-order WKB approach with Padé approximants, time-domain integration,  Frobenius, and pseudospectral methods. The Frobenius method, in particular, allows us to determine precise values of the frequencies, including the overtones. The two models differ in their choice of quantum parameter $\xi$, and we can distinguish them by their quasinormal spectra. In the first model, increasing the quantum parameter results in higher real oscillation frequencies and damping rates of the fundamental mode. In contrast, the second model shows a decrease in the oscillation frequency of the least-damped mode when the quantum parameter is introduced.  We have shown that, while the fundamental mode changes relatively gradually with the quantum parameter, the first few overtones deviate from their Schwarzschild limits at an increasing rate. This results in a qualitatively new behavior: the real parts of the frequencies of the first and higher overtones tend to zero as the quantum parameter increases. In addition to the branch of modes that are perturbative in the quantum parameter, we observe some non-perturbative modes at moderate values of the quantum parameter.
 Additionally, we have calculated the radii of the shadows cast by these black holes and discussed possible constraints based on observations of Sgt $A^*$. As a byproduct, we tested the method of calculating quasinormal modes of this kind based on a recent parametrization of effective potentials, and showed that while the parametrized formalism could be used for estimating the fundamental mode at small values of the coupling, its accuracy is highly dependent on the particular spacetime under consideration and is insufficient even for the lowest overtones.

\end{abstract}
\maketitle
\section{Introduction}

Efforts to find quantum corrections to the black hole metric involve various theoretical frameworks aimed at integrating quantum field theory with general relativity. Loop Quantum Gravity (LQG) modifies spacetime near singularities, providing non-singular black hole solutions \cite{Ashtekar:2005qt}. String theory introduces higher-order curvature corrections and interprets black hole entropy through microstates \cite{Strominger:1996sh}. Effective Field Theory (EFT) approaches add higher-derivative terms to the gravitational action, leading to modified metrics \cite{Donoghue:1994dn}. Non-commutative geometry proposes a fundamental length scale, smoothing out singularities \cite{Nicolini:2005vd}. The AdS/CFT correspondence translates quantum corrections from conformal field theory to modifications in the AdS black hole metric \cite{Maldacena:1997re}. Lastly, the asymptotic safety program uses renormalization group flow to derive consistent high-energy black hole solutions \cite{Niedermaier:2006wt}.

A particular approach to quantum corrections for the black hole spacetime we are interested about is related to  the  Hamiltonian constraints approach \cite{Thiemann:2007zz,Ashtekar:2004eh}.
The Hamiltonian constraints approach is a significant method in the quest for quantum gravity, particularly in the canonical quantization of general relativity. This approach involves reformulating Einstein's equations in terms of a Hamiltonian framework, where the dynamics of the gravitational field are governed by constraints. These constraints ensure the preservation of general covariance, meaning that the physical predictions do not depend on the choice of coordinates. In the context of black holes, the Hamiltonian constraints approach allows for the inclusion of quantum corrections by modifying the Hamiltonian to incorporate quantum effects.

Recently, a long-standing issue regarding general covariance in spherically symmetric gravity, which arises when canonical quantum gravity leads to a semiclassical model of black holes, was addressed in \cite{Zhang:2024khj}. This work proposed two black hole models that differ based on the choice of a quantum parameter.

The fundamental characteristic of black hole geometry is its spectrum of quasinormal modes \cite{Kokkotas:1999bd,Konoplya:2011qq}, which can be observed through gravitational wave interferometers \cite{LIGOScientific:2017vwq,LIGOScientific:2020zkf}. Future experiments promise to detect a much broader range of frequencies \cite{LISA:2022kgy}. While the fundamental mode primarily depends on the peak of the potential barrier, the first few overtones describe the geometry near the event horizon \cite{Konoplya:2023hqb,Konoplya:2022pbc}. In the time domain, these first overtones are crucial for describing the initial stage of the ringdown \cite{Giesler:2019uxc}. Observations in the gravitational spectrum can be complemented by electromagnetic observations, such as measuring the shadows cast by black holes \cite{EventHorizonTelescope:2019dse} and analyzing other optical phenomena \cite{Bambi:2015kza}.

While numerous studies have explored quasinormal modes of quantum-corrected black holes \cite{Bolokhov:2023bwm,Bolokhov:2023ruj,Skvortsova:2024atk,Dubinsky:2024fvi,Dubinsky:2024aeu,Bolokhov:2023ozp,Konoplya:2019xmn,Konoplya:2023ahd,Gong:2023ghh,Cruz:2020emz,Daghigh:2020fmw,2024arXiv240707892S,2024arXiv240719654Z,Calza:2024fzo,Calza:2024xdh}, the recent models produced within the effective Hamiltonian approach, maintaining general covariance, as presented in \cite{Zhang:2024khj}, have not been previously considered.

In this study, we investigate the quasinormal frequencies of the two black hole metrics derived in \cite{Zhang:2024khj} for scalar, electromagnetic, and effective axial gravitational perturbations. We employ four independent methods for these calculations: the higher-order WKB approach with Padé approximants, time-domain integration, the Frobenius method and the pseudospectral method. The Frobenius method, in particular, provides precise results, enabling us to determine accurate values for the overtones. Our findings reveal that while the fundamental mode changes relatively mildly upon the introduction of quantum corrections, the first few overtones deviate significantly, with their real oscillation frequencies tending towards zero. This behavior creates a distinctive "sound" of the event horizon deformed by quantum corrections. When the quantum parameter is not sufficiently small, we observe quasinormal modes that do not transition into the Schwarzschild modes as the quantum parameter approaches zero. In addition, we calculate the radius of the shadow cast by these two black holes.

We also used this study as an opportunity to test the parametrization of the effective potential suggested in~\cite{Cardoso_2019} as a tool for finding quasinormal modes. For this purpose, we have considered not only the two quantum-corrected black hole models developed in~\cite{Zhang:2024khj}, but also the black hole spacetime recently found as a result of quantum-corrected collapse in ~\cite{Lewandowski_2023}. While test field perturbations have been recently considered for this model in~\cite{Gong:2023ghh,Skvortsova:2024atk,Zinhailo:2024kbq,Luo:2024dxl}, no analysis of gravitational spectra has been done so far. In this way, we also complemented the existing works by studying the most valuable gravitational perturbations. It turns out that the parametrized approach~\cite{Cardoso_2019} can be used for calculating the fundamental quasinormal modes only, and even in that case, its accuracy is highly dependent on the model under consideration. For overtones, the relative error is usually of the same order as (or larger than) the effect (i.e., the deviation of the frequencies from their Schwarzschild values).

The structure of our work is as follows. In Section II, we introduce the black hole metrics and discuss the wave equations for scalar, electromagnetic, and effective gravitational perturbations. Section III reviews the four methods used for calculating the quasinormal frequencies. In Section IV, we present the numerical results for the quasinormal modes and provide analytical solutions in the eikonal regime. 
Section V is devoted to the discussion of the shadows cast by these black holes.
Finally, in the Conclusions, we discuss our findings and highlight some open problems.

\section{Black hole metric and wave-like equations}\label{sec:wavelike}

\subsection{The metric and the underlying theory}

The work of \cite{Zhang:2024khj} addresses the issue of maintaining covariance within the context of the spherically symmetric sector of vacuum gravity. By retaining the theory's kinematical variables and the classical form of the vector constraints, the study introduces an arbitrary effective Hamiltonian constraint, $H_{eff}$, along with a freely chosen function in constructing the effective metric. It is assumed, as in the classical theory, that a Dirac observable representing the black hole mass exists. Given these assumptions, the authors establish conditions on this observable and derive equations that ensure spacetime covariance. These conditions lead to relationships between the effective Hamiltonian, the Dirac observable for the black hole mass, and the free function. Solving these equations yields two families of effective Hamiltonian constraints, each parameterized by its own quantum parameter. Setting these parameters to zero recovers the classical constraints. Consequently, these effective Hamiltonian constraints produce two distinct quantum-corrected metrics, resulting in different spacetime structures.

The metric of the quantum-corrected black hole is given by the following line element
\begin{equation}\label{metric}
ds^2=-f(r)dt^2+\frac{1}{g(r)}dr^2+r^2(d\theta^2+\sin^2\theta d\phi^2),
\end{equation}
where for the first type black-hole model, the metric functions are
$$
\begin{array}{rcl}
\label{eq:metric_model_1}
f(r)&=&\displaystyle \left(1- \frac{2 M}{r}\right)\left[1+ \frac{\xi ^2}{r^2}\left(1-\frac{2 M}{r}\right)\right],\\
g(r)&=&f(r).\\
\end{array}
$$
Here $\xi $ is the quantum parameter, and $M$ is the ADM mass.

For the second black hole model, we have:
$$
\begin{array}{rcl}
\label{eq:metric_model_2}
f(r)&=&\displaystyle 1-\frac{2 M}{r},\\
g(r)&=&\displaystyle f(r) \left(1+\frac{\xi ^2}{r^2}f(r)\right).\\
\end{array}
$$
The advantage of this black hole metric is that it maintains the same relationship between the event horizon radius $r_{h}^+$ and the black hole mass as the classical solution, $r_{h}^+= 2M$. This consistency facilitates an easy comparison between the spectra of classical and quantum-corrected black holes.  

\subsection{Perturbation of test fields}
The general relativistic equations for the scalar ($\Phi$) and electromagnetic ($A_\mu$), can be written in the following form:
\begin{subequations}\label{coveqs}
\begin{eqnarray}\label{KGg}
\frac{1}{\sqrt{-g}}\partial_\mu \left(\sqrt{-g}g^{\mu \nu}\partial_\nu\Phi\right)&=&0,
\\\label{EmagEq}
\frac{1}{\sqrt{-g}}\partial_{\mu} \left(F_{\rho\sigma}g^{\rho \nu}g^{\sigma \mu}\sqrt{-g}\right)&=&0\,,
\end{eqnarray}
\end{subequations}
where $F_{\mu\nu}=\partial_\mu A_\nu-\partial_\nu A_\mu$ is the electromagnetic tensor.

After separation of the variables in the background (\ref{metric}) the above equations (\ref{coveqs}) take the Schrödinger wavelike form \cite{Kokkotas:1999bd,Berti:2009kk,Konoplya:2011qq}:
\begin{equation}\label{wave-equation}
\dfrac{d^2 \Psi}{dr_*^2}+(\omega^2-V(r))\Psi=0,
\end{equation}
where the ``tortoise coordinate'' $r_*$ is defined as follows:
\begin{equation}\label{tortoise}
dr_*\equiv\frac{dr}{\sqrt{f(r) g(r)}}.
\end{equation}

The effective potentials for the scalar ($s=0$) and electromagnetic ($s=1$) fields have the form
\begin{equation}\label{potentialScalar}
V(r)=f(r)\frac{\ell(\ell+1)}{r^2}+\frac{1-s}{r}\cdot\frac{d^2 r}{dr_*^2},
\end{equation}
where $\ell=s, s+1, s+2, \ldots$ are the multipole numbers.

\begin{figure}
\resizebox{\linewidth}{!}{\includegraphics{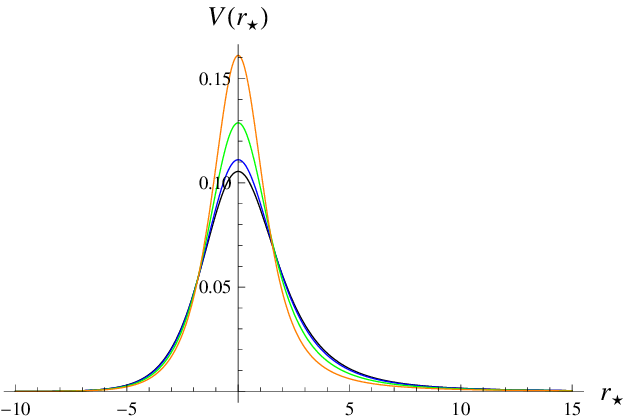}}
\caption{Effective potential as a function of the tortoise coordinate of the $\ell=0$ scalar field perturbations of the first black hole model  ($M=1/2$): $\xi=0$ (black), $\xi=0.4$ (blue), $\xi=0.8$ (green), $\xi=1.2$ (orange).}\label{fig:potentials1}
\end{figure}

\begin{figure}
\resizebox{\linewidth}{!}{\includegraphics{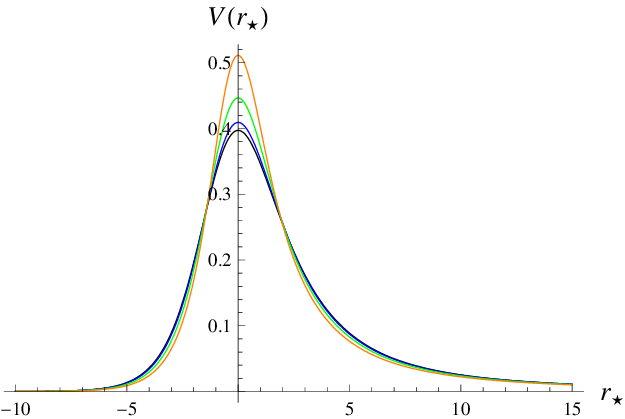}}
\caption{Effective potential as a function of the tortoise coordinate of the $\ell=1$ scalar field perturbations of the first black hole model ($M=1/2$): $\xi=0$ (black), $\xi=0.4$ (blue), $\xi=0.8$ (green), $\xi=1.2$ (orange).}\label{fig:potentials2}
\end{figure}

\begin{figure}
\resizebox{\linewidth}{!}{\includegraphics{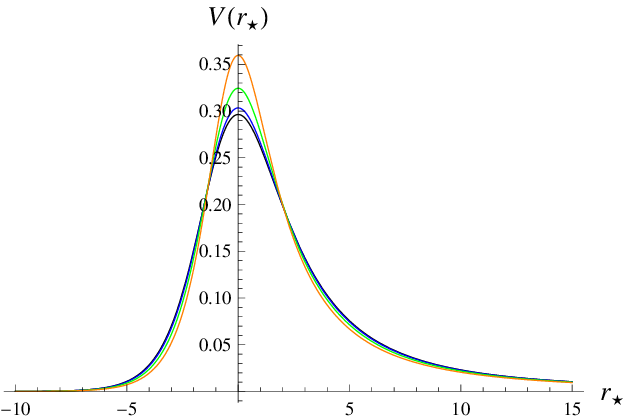}}
\caption{Effective potential as a function of the tortoise coordinate of the $\ell=1$ electromagnetic field perturbations of the first black hole model ($M=1/2$): $\xi=0$ (black), $\xi=0.4$ (blue), $\xi=0.8$ (green), $\xi=1.2$ (orange).}\label{fig:potentials3}
\end{figure}

The effective potentials for the above fields are shown in figs. \ref{fig:potentials1}--\ref{fig:potentials3}. They are positive definite, which guarantees stability for these perturbations. 

\subsection{Perturbations of gravitational field}

\begin{figure}
\resizebox{\linewidth}{!}{\includegraphics{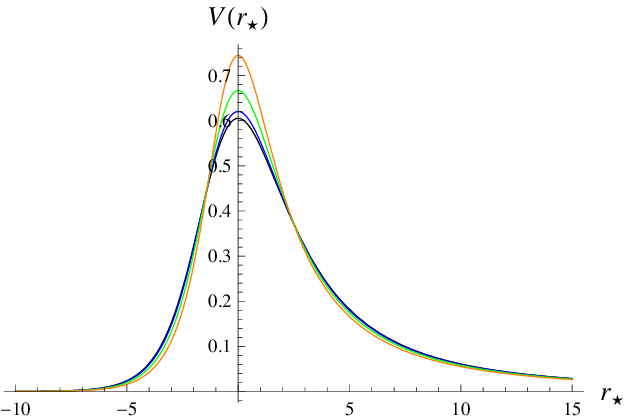}}
\caption{Effective potential as a function of the tortoise coordinate of the $\ell=2$ axial gravitational perturbations of the first black hole model ($M=1/2$): $\xi=0$ (black), $\xi=0.4$ (blue), $\xi=0.8$ (green), $\xi=1.2$ (orange).}\label{fig:potentials4}
\end{figure}

The problem of gravitational perturbations is more complex in this case because the metric is derived through an effective approach using Hamiltonian constraints, rather than being an exact solution of the Einstein equations with quantum corrections. Consequently, a rigorous analysis of gravitational perturbations is challenging to achieve. However, as demonstrated by Ashtekar, Olmedo, and Singh \cite{Ashtekar:2018lag,Ashtekar:2018cay}, quantum corrections can be effectively modeled as contributions from an anisotropic fluid's energy-momentum tensor within the framework of Einsteinian gravity. This allows for the study of perturbations in such a system. Following the work of Bouhmadi-López et al. \cite{Bouhmadi-Lopez:2020oia}, axial perturbations can be analyzed under the assumption that perturbations in the direction of the anisotropy are negligible in the axial sector of gravitational perturbations.

There are several similar instances where certain perturbations are considered relatively minor and thus neglected \cite{Berti:2003yr,Kokkotas:1993ef,Konoplya:2006ar}. While this approach may overlook some significant features of the gravitational spectrum, it serves as a reasonable approximation, especially when the black hole geometry deviates only slightly from the classical Schwarzschild limit. This approximation aligns with the concept of perturbative quantum corrections, which are expected to be relatively small.

The metrics of the axial gravitational perturbations $h_{\mu\nu}$ in the Regge-Wheeler gauge \cite{Regge:1957td} take the following form
\begin{widetext}
\begin{eqnarray}
h^{axial}_{\mu \nu}= \left[
 \begin{array}{cccc}
 0 & 0 &0 & h_0(t,r)
\\ 0 & 0 &0 & h_1(t,r)
\\ 0 & 0 &0 & 0
\\ h_0(t,r) & h_1(t,r) &0 &0
\end{array}\right]
\left(\sin\theta\frac{\partial}{\partial\theta}\right)
P_{\ell}(\cos\theta)\,, \label{pert_axial}
\end{eqnarray}
where $h_0(t,r)$ and $h_1(t,r)$ are two unknown functions, and $P_{\ell}(x)$  is the Legendre polynomial. 
\end{widetext}
The solutions can be considered as solutions of the Einstein equations with some anisotropic fluid with
\begin{equation}
    T_{\mu\nu}=(\rho+p_t)u_{\mu\nu}+g_{\mu\nu}p_t+(p_r-p_t)s_{\mu}s_{\nu},
\end{equation}
where  $\rho$, $p_r$, $p_t$ are the fluid density, radial, and tangential pressure, respectively. The fluid velocity $u_{\mu}$, and radial space-like unit vector $s_{\mu}$, are given by 
\begin{equation}
    u_{\mu}=(\sqrt{f(r)},0,0,0),\,s_{\mu}=(0,1/\sqrt{g(r)},0,0).
\end{equation}
They satisfy
\eq{
u_{\mu}u^{\mu}=-1,\,s_{\mu}s^{\mu}=1,\,u_{\mu}s^{\mu}=0.
}
The quantities $\rho$, $p_r$, $p_t$ transform as scalars  relatively the rotation group on the two dimensional sphere, so their axial perturbations are zero.  
For $u_{\mu}$ and $s_{\mu}$ vectors we have nonzero  perturbations components 
\eq{
\delta u_{\phi}=-i\omega U(r)e^{-i\omega t}\sin\theta\partial_{\theta}P_{\ell}(\cos\theta) 
}
\eq{
\delta s_{\phi}=-S(r)e^{-i\omega t}\sin\theta\partial_{\theta}P_{\ell}(\cos\theta),
}
Further, we assume that there are no perturbations in the anisotropy direction, i.e. $\delta s_{\mu}=0$.

\begin{figure}
\resizebox{\linewidth}{!}{\includegraphics{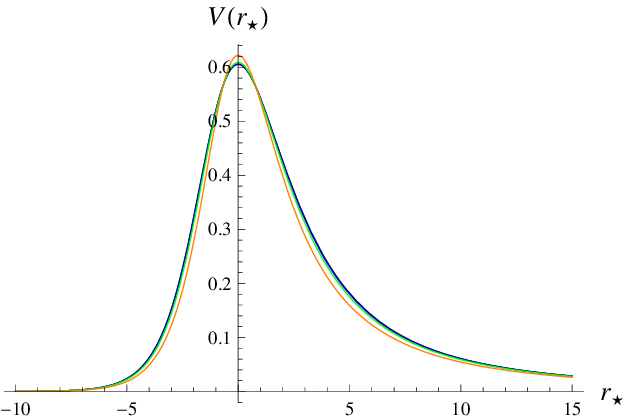}}
\caption{Effective potential as a function of the tortoise coordinate of the $\ell=2$ axial gravitational perturbations of the second black hole model ($M=1/2$): $\xi=0$ (black), $\xi=0.4$ (blue), $\xi=0.8$ (green), $\xi=1.6$ (orange).}\label{fig:potentials5}
\end{figure}

From $\nabla_{\mu}T^{\mu r}=0$, we obtain that $\delta u_{\phi}=0$.
After substitution into the Einstein equations
we obtain
\begin{align}
        & h_1(r) \left(r^2 \omega^2 - (\ell - 1)(\ell + 2) f(r)\right) \nonumber \\
        &\quad - i r^2 \omega h_0'(r) + 2 i r \omega h_0(r) = 0,
\end{align}
\eq{
f(r) \left(\frac{h_1(r) f'(r)}{f(r)}+2 h_1'(r)\right)+\frac{2 i
   \omega  h_0(r)}{g(r)}=0,
}
After simple algebra and introducing new variables
\eq{h_1=\frac{r}{\sqrt{f(r)g(r)}}\Psi,\quad dr_*=\frac{dr}{\sqrt{f(r)g(r)}},}
we obtain
\eq{
\frac{d^2}{dr_*^2}\Psi+\left[\omega^2-V_{ax}(r)\right]=0,}
\eq{V_{ax}=f(r)\left(\frac{2g(r)}{r^2}-\frac{(fg)'}{2rf}+\frac{(\ell+2)(\ell-1)}{r^2}
\right).
}

A similar approach  was used in \cite{Bronnikov:2012ch} for studying perturbations of a black hole spacetime with a scalar field, where the scalar field perturbations decouple from the axial gravitational perturbations (see also \cite{Chen:2019iuo}).

The effective potential for axial gravitational perturbations is shown in figs. \ref{fig:potentials4} and \ref{fig:potentials5}. There one can see that the effective potentials are positive definite, which guarantees the stability of the black hole model at least within the framework of the considered effective types of perturbations.

\section{Methods}\label{sec:methods}

Here we briefly review the four methods used for the calculations of quasinormal modes: the  WKB method, time-domain integration,  the Frobenius method, and the pseudospectral method. The Frobenius method is based on a converging procedure and, therefore, provides precise values of quasinormal frequencies.

By definition, quasinormal modes satisfy the following boundary conditions,
\begin{equation}\label{boundaryconditions}
\Psi(r_*\to\pm\infty)\propto e^{\pm\imo \omega r_*},
\end{equation}
which are requirement of the purely ingoing waves at the event horizon ($r_*\to-\infty$) and purely outgoing wave at spatial infinity ($r_*\to\infty$).

\subsection{WKB method}

The WKB approach is an effective and apparently the most economic way to find quasinormal modes with $\ell \geq n$. 
It consists of matching of the asymptotic WKB solutions with the Taylor expansion of the wave function around the maximum of the potential barrier. The general WKB formula can be written in the form of expansion around the eikonal limit ( $\ell \gg n$) \cite{Konoplya:2019hlu}:
\begin{eqnarray}\label{WKBformula-spherical}
\omega^2&=&V_0+A_2(\K^2)+A_4(\K^2)+A_6(\K^2)+\ldots\\\nonumber&-&\imo \K\sqrt{-2V_2}\left(1+A_3(\K^2)+A_5(\K^2)+A_7(\K^2)\ldots\right),
\end{eqnarray}
and the matching conditions under the assumptions of the quasinormal modes boundary conditions produce
\begin{equation}
\K=n+\frac{1}{2}, \quad n=0,1,2,\ldots,
\end{equation}
where $n$ is the overtone number, and $V_i$ is the value of the $i-th$ derivative of effective potential at its maximum relatively the tortoise coordinate. The functions $A_i$ for $i=2, 3, 4, \ldots$ are $i-th$ WKB order correction terms to the eikonal limit, which depends on $\K$ and derivatives of the potential in its maximum up to the order $2i$. The explicit forms of $A_i$ at various $i$ can be found in \cite{Iyer:1986np,Konoplya:2003ii,Konoplya:2004ip,Matyjasek:2017psv}. In the present paper we used the 6th order WKB method \cite{Konoplya:2003ii} with Pade approximants  \cite{Matyjasek:2017psv}. For the Pade approximants we used $\tilde{m}=4$ as defined in \cite{Konoplya:2019hlu}.

\subsection{Time-domain integration}
Using integration in the time domain at a fixed value of the radial coordinate, one can see the evolution of the wave function. 
For such integration we use the Gundlach-Price-Pullin discretization scheme \cite{Gundlach:1993tp}
\begin{eqnarray}
\Psi\left(N\right)&=&\Psi\left(W\right)+\Psi\left(E\right)-\Psi\left(S\right)\nonumber\\
&&- \Delta^2V\left(S\right)\frac{\Psi\left(W\right)+\Psi\left(E\right)}{4}+{\cal O}\left(\Delta^4\right),\label{Discretization}
\end{eqnarray}
where the points are defined as follows: $N\equiv\left(u+\Delta,v+\Delta\right)$, $W\equiv\left(u+\Delta,v\right)$, $E\equiv\left(u,v+\Delta\right)$, and $S\equiv\left(u,v\right)$. 

Then, in order to extract the values of frequencies from the time-domain profile, we use the Prony method, which consists in fitting of the profile data by a sum of exponents with some weights:
\begin{equation}\label{damping-exponents}
\Psi(t)\simeq\sum_{i=1}^pC_ie^{-i\omega_i t}.
\end{equation}
Then, assuming that the ringing starts at some time, we can find the quasinormal frequencies. 

\subsection{Frobenius method}

The Frobenius method for solutions of differential equations has been well-known for a long time and was applied for the first time by  Leaver \citep{Leaver:1985ax,Leaver:1986gd} to the problem of finding of quasinormal modes of black holes. The method is based on expansion into converging series and therefore it gives the frequencies with any desired precision. For quicker convergence we use the Nollert technique  \citep{Nollert:1993zz} in its general form developed in \citep{Zhidenko:2006rs}.

The wave-like equation has regular singular points at $r=0$, at the inner and outer horizons $r=r_h^{-}$, $r=r_h^{+}$, and an irregular singular point at $r=\infty$.
We introduce a new radial function $P(r, \omega)$, 
\begin{equation}\label{reg}
\Psi(r)= P (r, \omega) y(r),
\end{equation}
such that the factor $P(r, \omega)$  makes $y(r)$ regular in the range $r_h^{+}\leq r$ when the quasinormal modes boundary conditions are satisfied.
Then, expanding $y(r)$ as follows 
\begin{equation}\label{Frobenius}
y(r)=\sum_{k=0}^{\infty}a_k\left(\frac{r-r_h^{+}}{r-r_h^{-}}\right)^k,
\end{equation}
with 
\eq{
P (r, \omega)=\left(\frac{r-r_h^{+}}{r-r_h^{-}}\right)^{-\frac{i\omega}{f'(r_h^{+})}}(r-r_h^{-})^{2ir_h^{+}\omega }e^{i\omega r},
}
and using  Gaussian eliminations, we reduce finding of $\omega$ to the problem of numerical solution of a non-algebraic equation via  the FindRoot command in {\it Mathematica}. If even after the above procedure for a chosen $P(r, \omega)$ at some $\omega$, the singular points appear between the event horizon and infinity, we use  integration though a sequence of positive real midpoints as suggested by \cite{Rostworowski:2006bp}.

\subsection{Pseudospectral method} 
The Chebyshev pseudospectral method is a powerful technique for solving differential equations. The method is based on discretizing the unknown function over a grid of collocation points, typically corresponding to the roots of Chebyshev polynomials. By transforming the differential equations into a system of algebraic equations at these collocation points and solving them, we can determine the function's values at grid points.

To ensure the correct boundary conditions corresponding to quasinormal modes, we make a substitution in the master equation (\ref{wave-equation})
\eq{
\Psi(r)=r^{2i r_h^+\omega}(r-r_h^+)^{-i r_h^+\omega}e^{i\omega r}y(r),
}
which leads to $y\sim const$ at the outer black hole horizon and at spatial infinity.

Then, we  compactify our semi-interval $[r_h,\infty)$ to interval $[0,1]$ by introducing a new variable
\eq{
r=\frac{r_h}{1-u},
}
Altogether, these substitutions lead to an equation of the form
\eq{\label{eq:reg}
A_2(u)y''(u)+A_1(u)y'(u)+A_0(u)y(u)=0,
}
where $A_i(u)=A_i(u,\omega,\omega^2)$, $i=0,1,2$. At the next step, we discretize equation (\ref{eq:reg}) on the Chebyshev-Lobatto grid, which is defined as
\eq{u_j=\frac{1}{2}\left(1-\cos\left[\frac{\pi j}{N}\right]\right),~~j=0,1...N.}
This results in a matrix equation
\eq{\label{eq:discr_eq}
\left(\tilde{M}_0+\tilde{M}_1
\omega+\tilde{M}_2\,\omega^2\right)\tilde{y}=0,
}
where $\tilde{y}$  is the vector of the unknown function's values, $\tilde{M}_i$   are the numerical matrices of discretized coefficients at the collocation grid points.

We can linearize our quadratic eigenproblem (\ref{eq:discr_eq}) in the following way
\eq{\label{eq:LinEig_eq}
\left(M_0+M_1\omega\right)\tilde{\psi}=\mathbb{0},}
where  
\begin{equation}
M_0=
\begin{pmatrix}
\tilde{M}_0 & \tilde{M}_1 \\
\mathbb{0} & \mathbb{1}
\end{pmatrix},\,
M_1=
\begin{pmatrix}
\mathbb{0} & \tilde{M}_2 \\
-\mathbb{1} & \mathbb{0}
\end{pmatrix},\,
\tilde{\psi}=
\begin{pmatrix}
\tilde{y}\\
\omega\tilde{y}
\end{pmatrix}.
\end{equation}

Then QNMs spectrum can be
found by solving generalized eigenvalue problem (\ref{eq:LinEig_eq}) via  the Eigenvalues command in {\it Mathematica}. To avoid spurious eigenvalues we perform  the calculations on two grids of different sizes and select only overlapping values \cite{boyd2013chebyshev}.
\begin{figure*}
\resizebox{\linewidth}{!}{\includegraphics{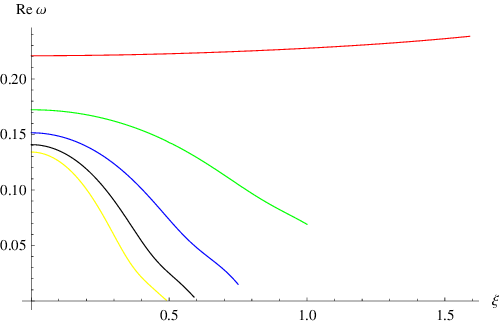}~~~\includegraphics{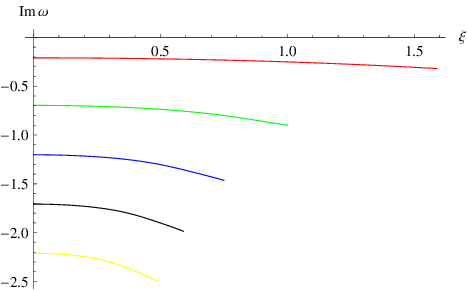}}
\caption{The fundamental mode and the first four overtones as a function of $\xi$ for the first black-hole model; $\ell=0$ scalar field perturbations.}\label{fig:BH1L0s0}
\end{figure*}

\begin{figure*}
\resizebox{\linewidth}{!}{\includegraphics{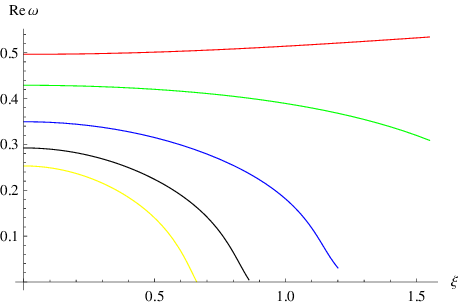}~~~\includegraphics{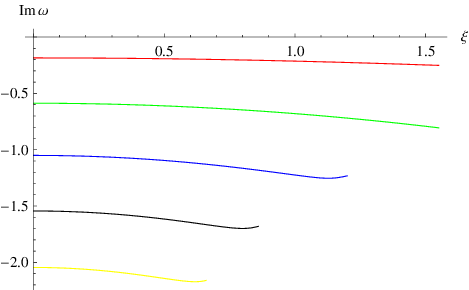}}
\caption{The fundamental mode and the first four overtones as a function of $\xi$ for the first black-hole model; $\ell=1$ electromagnetic field perturbations.}\label{fig:BH1L1s1}
\end{figure*}

\begin{figure*}
\resizebox{\linewidth}{!}{\includegraphics{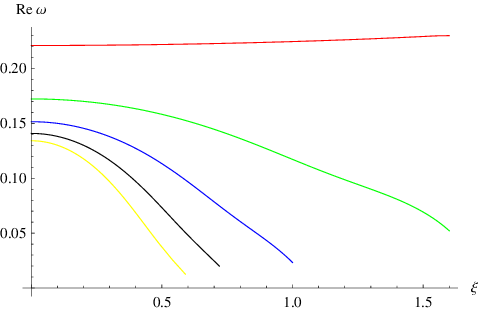}~~~\includegraphics{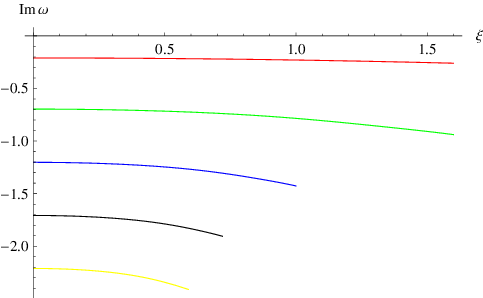}}
\caption{The fundamental mode and the first four overtones as a function of $\xi$ for the second black-hole model; $\ell=0$ scalar field perturbations.}\label{fig:BH1L0s0}
\end{figure*}

\begin{figure*}
\resizebox{\linewidth}{!}{\includegraphics{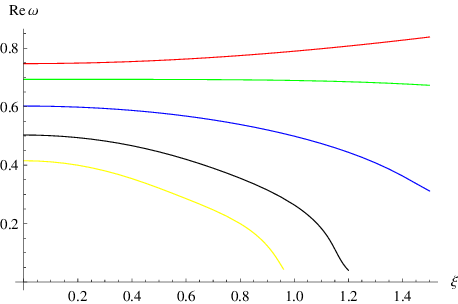}~~~\includegraphics{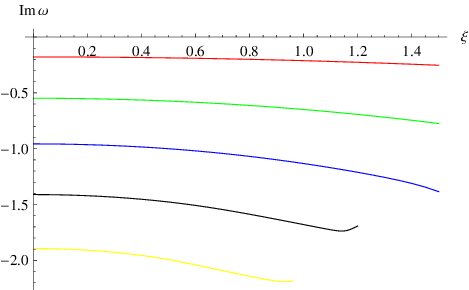}}
\caption{The fundamental mode and the first four overtones as a function of $\xi$ for the first black-hole model; $\ell=2$ axial gravitational perturbations.}\label{fig:BH1L2s2}
\end{figure*}

\begin{figure*}
\resizebox{\linewidth}{!}{\includegraphics{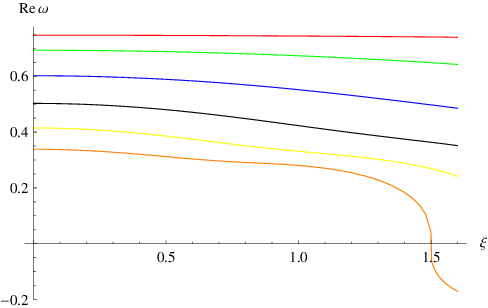}~~~\includegraphics{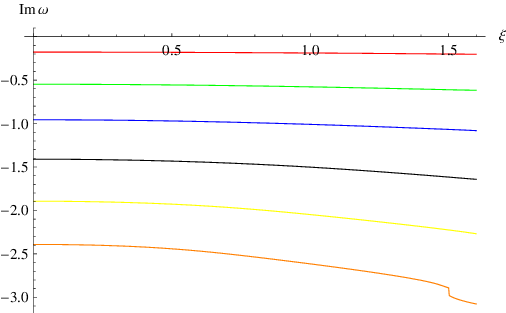}}
\caption{The fundamental mode and the first five overtones as a function of $\xi$ for the second black hole model; $\ell=2$ axial gravitational perturbations.}\label{fig:NNBH2L2s2}
\end{figure*}

\begin{figure*}
\resizebox{\linewidth}{!}{\includegraphics{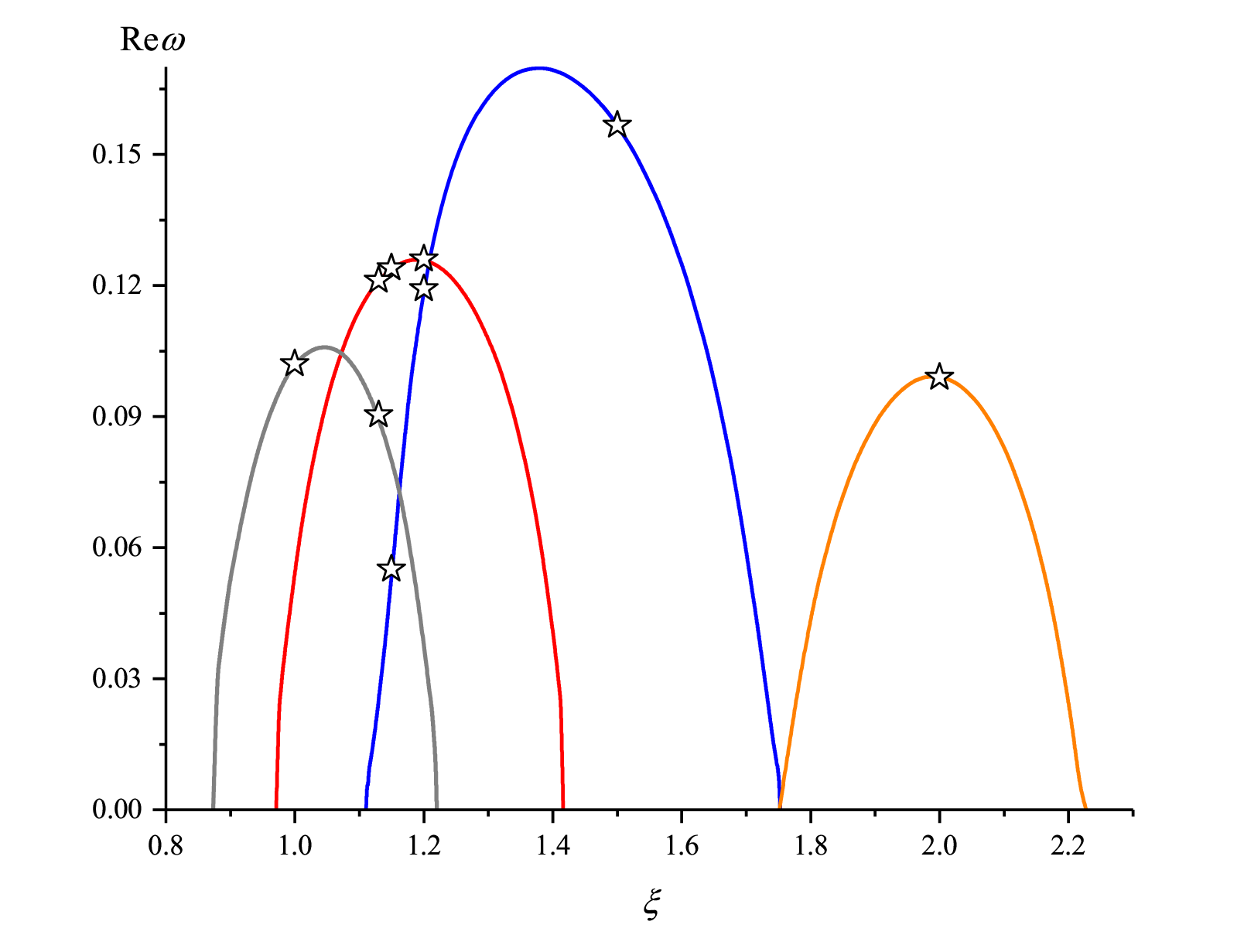}~~~\includegraphics{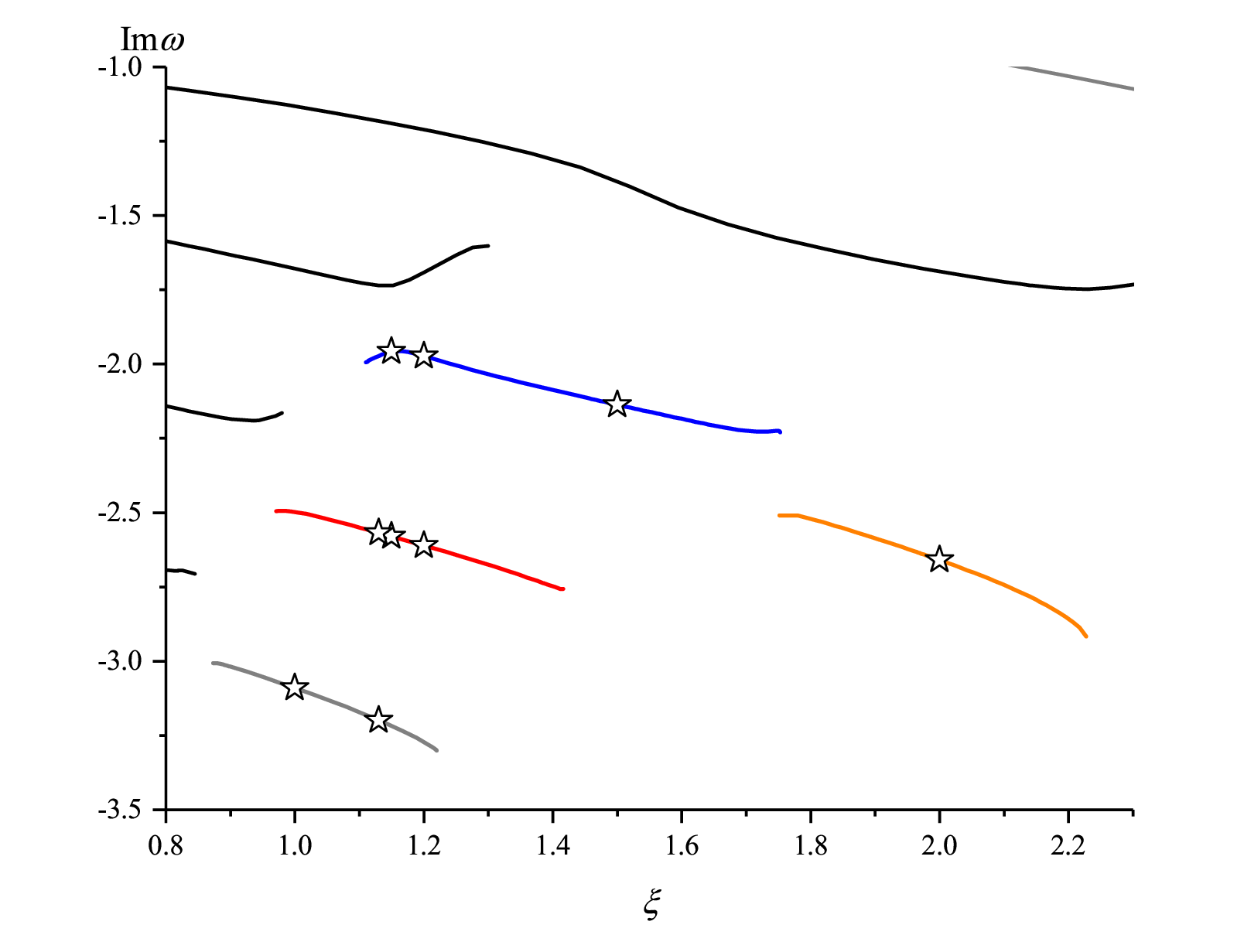}}
\caption{The first non-perturbative modes as a function of $\xi$ for the first black hole model; $\ell=2$ axial gravitational perturbations. The star symbols correspond to the values from the Table. \ref{tableXII}}\label{fig:Non-Perturbative}
\end{figure*}
\begin{figure}
\resizebox{\linewidth}{!}{\includegraphics{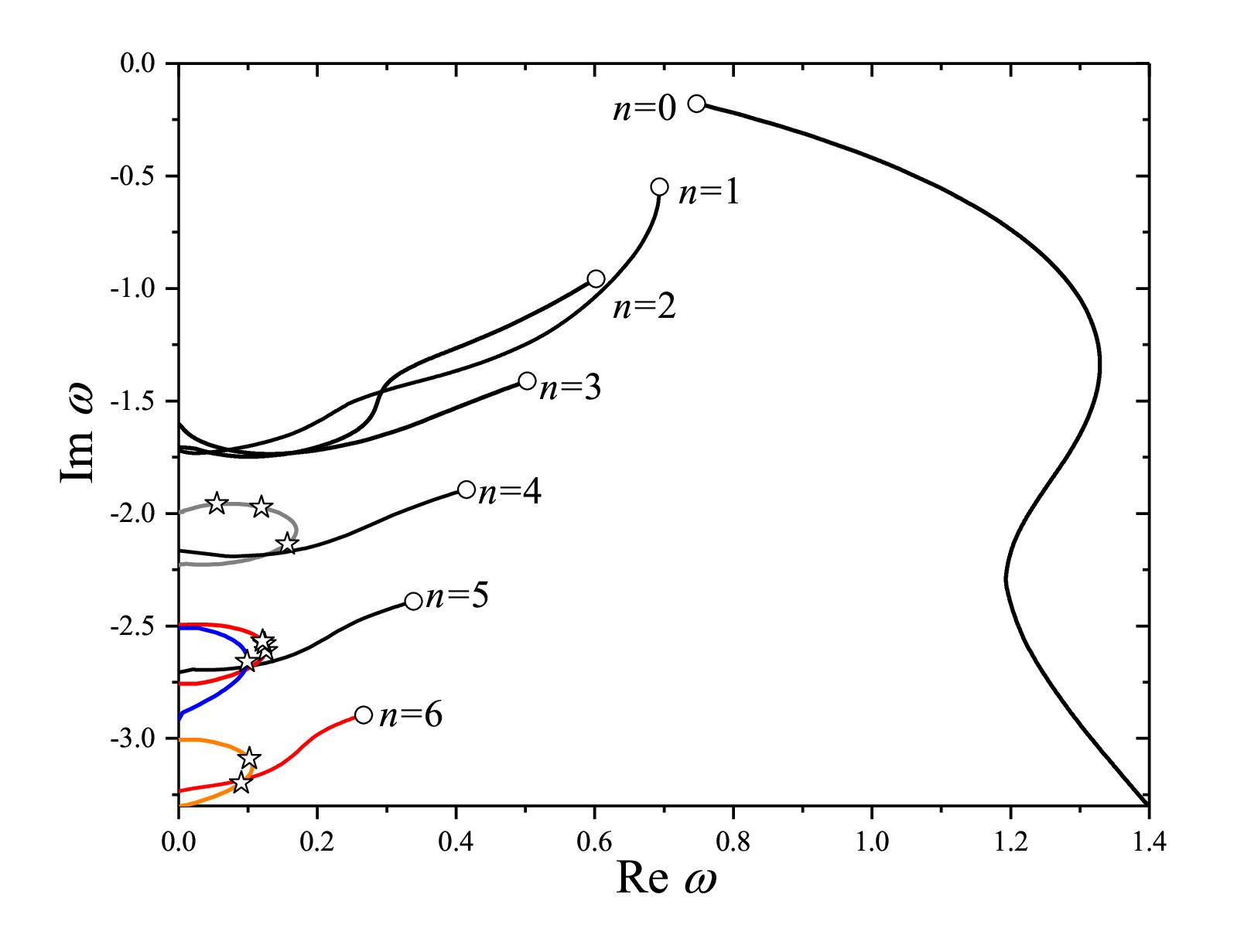}}
\caption{The complex $\omega$ plane of QNMs. : The black lines represent the perturbative branches, while the colored lines correspond to the non-perturbative branches for the first black hole model; $\ell=2$ axial gravitational perturbations.}\label{fig:Non-Perturbative}
\end{figure}
\begin{figure}
\resizebox{\linewidth}{!}{\includegraphics{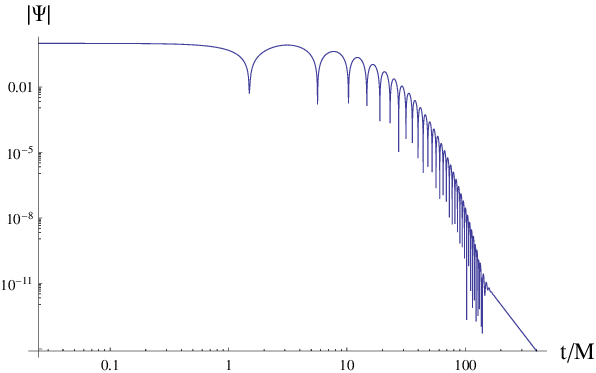}}
\caption{Time-domain profile for axial gravitational perturbations of the first black hole model, $\ell=2$, $M=1/2$, $\xi=0.3$.}\label{fig:TD1}
\end{figure}

\begin{table*}
\begin{tabular}{c c c c}
\hline
\hline
$\xi $ & Prony fit & WKB6 Padé &  Frobenius\\
\hline
$0$ & $0.218387-0.209629 i$ & $0.221584-0.209367 i$ &     $0.220910 - 0.209791 i$\\
$0.2$ & $0.218610-0.211268 i$ & $0.221098-0.211264 i$ &  $0.221198 - 0.211371 i$\\
$0.4$ & $0.219275-0.216279 i$ & $0.219706-0.215676 i$ &   $0.222049 - 0.216145 i$\\
$0.6$ & $0.220437-0.224933 i$ & $0.221115-0.224992 i$ &  $0.223427 - 0.224215 i$\\
$0.8$ & $0.222389-0.237582 i$ & $0.226484-0.243023 i$ &     $0.225296 - 0.235748 i$\\
$1.$ & $0.225775-0.254226 i$ & $0.254696-0.241289 i$ &   $0.227645 - 0.250970 i$\\
$1.2$ & $0.230904-0.274022 i$ & $0.235955-0.250742 i$ &    $0.230537 - 0.270125 i$ \\
$1.4$ & $0.236851-0.296233 i$ & $0.234166-0.275977 i$ &    $0.234136 - 0.293402 i$\\
\hline
\hline
\end{tabular}
\caption{Comparison of the scalar ($s=0$) quasinormal frequencies for the first BH model obtained by the time-domain integration and the 6th order WKB approach with Padé approximants and the Frobenius method for $\ell=0$ ($M=1/2$).}\label{tableI}
\end{table*}
\begin{table*}
\begin{tabular}{c c c c c}
\hline
\hline
$\xi $ & Prony fit & WKB6 Padé & rel. diff $Re (\omega)$ & rel. diff $Im (\omega)$  \\
\hline
$0$ & $0.585884-0.195294 i$ & $0.585859-0.195325 i$ & $0.00433\%$ & $0.0160\%$  \\
$0.2$ & $0.587292-0.196837 i$ & $0.587283-0.196879 i$ & $0.00151\%$ & $0.0215\%$ \\
$0.4$ & $0.591486-0.201473 i$ & $0.591499-0.201498 i$ & $0.00228\%$ & $0.0128\%$ \\
$0.6$ & $0.598390-0.209219 i$ & $0.598435-0.209223 i$ & $0.00759\%$ & $0.00208\%$ \\
$0.8$ & $0.607881-0.220103 i$ & $0.607916-0.220096 i$ & $0.00578\%$ & $0.00340\%$ \\
$1.$ & $0.619800-0.234158 i$ & $0.619845-0.234331 i$ & $0.00733\%$ & $0.0737\%$ \\
$1.2$ & $0.633960-0.251424 i$ & $0.633937-0.251460 i$ & $0.00355\%$ & $0.0143\%$  \\
$1.4$ & $0.650159-0.271950 i$ & $0.650062-0.272029 i$ & $0.0150\%$ & $0.0291\%$  \\
\hline
\hline
\end{tabular}
\caption{Comparison of the scalar ($s=0$) quasinormal frequencies for the first BH model obtained by the time-domain integration and the 6th order WKB approach with Padé approximants  $\ell=1$ ($M=1/2$).}\label{tableII}
\end{table*}
\begin{table*}
\begin{tabular}{c c c c}
\hline
\hline
$\xi $ & Prony fit & WKB6 Padé & Frobenius   \\
\hline
$0$ & $0.496524-0.184939 i$ & $0.496509-0.184993 i$ &   $0.496527 - 0.184975 i$ \\
$0.2$ & $0.497298-0.186108 i$ & $0.497313-0.186159 i$ & $0.497300 - 0.186143 i$ \\
$0.4$ & $0.499584-0.189603 i$ & $0.499662-0.189645 i$ & $0.499586 - 0.189633 i$ \\
$0.6$ & $0.503286-0.195384 i$ & $0.503416-0.195413 i$ & $0.503285 - 0.195407 i$ \\
$0.8$ & $0.508246-0.203385 i$ & $0.508395-0.203405 i$ & $0.508242 - 0.203402 i$ \\
$1.$ & $0.514264-0.213517 i$ & $0.514395-0.213536 i$ &  $0.514255 - 0.213530 i$ \\
$1.2$ & $0.521106-0.225669 i$ & $0.521194-0.225712 i$ & $0.521091 - 0.225683 i$ \\
$1.4$ & $0.528519-0.239711 i$ & $0.528569-0.239834 i$ & $0.528498 - 0.239730 i$ \\
\hline
\hline
\end{tabular}
\caption{Comparison of the electormagnetic ($s=1$) quasinormal frequencies for the first black hole model obtained by the time-domain integration and the 6th order WKB approach with Padé approximants and Frobenius method for $\ell=1$ ($M=1/2$).}\label{tableIII}
\end{table*}
\begin{table*}
\begin{tabular}{c c c c}
\hline
\hline
$\xi $ & Prony fit & WKB6 Padé & Frobenius   \\
\hline
$0$ & $0.218387-0.209629 i$ & $0.221584-0.209367 i$   & $0.220910 - 0.209791 i$ \\
$0.2$ & $0.218499-0.210446 i$ & $0.221485-0.210223 i$ & $0.221054 - 0.210580 i$ \\
$0.4$ & $0.218834-0.212908 i$ & $0.219813-0.212868 i$ & $0.221486 - 0.212943 i$ \\
$0.6$ & $0.219402-0.217052 i$ & $0.219874-0.216627 i$ & $0.222202 - 0.216876 i$ \\
$0.8$ & $0.220242-0.222932 i$ & $0.221167-0.223578 i$ & $0.223198 - 0.222368 i$ \\
$1.$ & $0.221461-0.230599 i$ & $0.225523-0.235366 i$ &  $0.224481 - 0.229407 i$ \\
$1.2$ & $0.223260-0.240035 i$ & $0.248042-0.240253 i$ & $0.226069 - 0.237973 i$ \\
$1.4$ & $0.225892-0.251044 i$ & $0.237632-0.229887 i$ & $0.227999 - 0.248035 i$ \\
\hline
\hline
\end{tabular}
\caption{Comparison of the quasinormal frequencies for the second BH model obtained by the time-domain integration and the 6th order WKB approach with Padé approximants for  $s=0$, $\ell=0$ ($M=1/2$).}\label{tableIV}
\end{table*}
\begin{table*}
\begin{tabular}{c c c c c}
\hline
\hline
$\xi $ & Prony fit & WKB6 Padé & rel. diff $Re (\omega)$ & rel. diff $Im (\omega)$  \\
\hline
$0$ & $0.585884-0.195294 i$ & $0.585859-0.195325 i$ & $0.00433\%$ & $0.0160\%$\\
$0.2$ & $0.585931-0.195908 i$ & $0.585917-0.195954 i$ & $0.00227\%$ & $0.0233\%$\\
$0.4$ & $0.586068-0.197743 i$ & $0.586066-0.197780 i$ & $0.00033\%$ & $0.0185\%$\\
$0.6$ & $0.586294-0.200778 i$ & $0.586314-0.200800 i$ & $0.00342\%$ & $0.0109\%$\\
$0.8$ & $0.586605-0.204978 i$ & $0.586647-0.204985 i$ & $0.00716\%$ & $0.00333\%$\\
$1.$ & $0.586997-0.210299 i$ & $0.587029-0.210295 i$ & $0.00542\%$ & $0.00186\%$\\
$1.2$ & $0.587463-0.216690 i$ & $0.587328-0.216803 i$ & $0.0230\%$ & $0.0519\%$\\
$1.4$ & $0.588000-0.224098 i$ & $0.588225-0.224214 i$ & $0.0383\%$ & $0.0516\%$\\
\hline
\hline
\end{tabular}
\caption{Comparison of the quasinormal frequencies for the second BH model obtained by the time-domain integration and the 6th order WKB approach with Padé approximants for $s=0$, $\ell=1$ ($M=1/2$).}\label{tableV}
\end{table*}
\begin{table*}
\begin{tabular}{c c c c c}
\hline
\hline
$\xi $ & Prony fit & WKB6 Padé & rel. diff $Re (\omega)$ & rel. diff $Im (\omega)$  \\
\hline
$0$ & $0.496524-0.184939 i$ & $0.496509-0.184993 i$ & $0.00310\%$ & $0.0295\%$\\
$0.2$ & $0.496186-0.185294 i$ & $0.496183-0.185348 i$ & $0.00066\%$ & $0.0289\%$\\
$0.4$ & $0.495174-0.186348 i$ & $0.495205-0.186400 i$ & $0.00620\%$ & $0.0281\%$\\
$0.6$ & $0.493501-0.188066 i$ & $0.493583-0.188122 i$ & $0.0166\%$ & $0.0299\%$\\
$0.8$ & $0.491187-0.190398 i$ & $0.491330-0.190469 i$ & $0.0291\%$ & $0.0376\%$\\
$1.$ & $0.488257-0.193274 i$ & $0.488464-0.193380 i$ & $0.0424\%$ & $0.0545\%$\\
$1.2$ & $0.484743-0.196619 i$ & $0.485010-0.196784 i$ & $0.0551\%$ & $0.0838\%$\\
$1.4$ & $0.480680-0.200350 i$ & $0.480996-0.200607 i$ & $0.0657\%$ & $0.128\%$\\
\hline
\hline
\end{tabular}
\caption{Comparison of the quasinormal frequencies for the second BH model obtained by the time-domain integration and the 6th order WKB approach with Padé approximants for  $s=1$, $\ell=1$ ($M=1/2$).}\label{tableVI}
\end{table*}
\begin{table*}
\begin{tabular}{c c c c}
\hline
\hline
$\xi $ & Prony fit & WKB6 Padé & Frobenius   \\
\hline
$0$ & $0.747356-0.177914 i$ & $0.747238-0.177857 i$ &   $0.747343 - 0.177925 i$ \\
$0.2$ & $0.749129-0.179257 i$ & $0.749131-0.179207 i$ & $0.749116 - 0.179267 i$ \\
$0.4$ & $0.754411-0.183281 i$ & $0.754477-0.183222 i$ & $0.754397 - 0.183291 i$ \\
$0.6$ & $0.763091-0.189975 i$ & $0.763323-0.189879 i$ & $0.763076 - 0.189986 i$ \\
$0.8$ & $0.774994-0.199321 i$ & $0.775636-0.199279 i$ & $0.774977 - 0.199333 i$ \\
$1.$ &  $0.789893-0.211295 i$ & $0.791334-0.211781 i$ & $0.789874 - 0.211309 i$ \\
$1.2$ & $0.807521-0.225870 i$ & $0.809481-0.227926 i$ & $0.807499 - 0.225885 i$ \\
$1.4$ & $0.827586-0.243013 i$ & $0.828953-0.246556 i$ & $0.827562 - 0.243031 i$ \\
$1.6$ & $0.849785-0.262694 i$ & $0.850556-0.266906 i$ & $0.849757-0.262714 i$ \\
\hline
\hline
\end{tabular}
\caption{Comparison of the quasinormal frequencies for axial gravitational perturbations ($s=2$) of the first black hole model obtained by the time-domain integration and the 6th order WKB approach with Padé approximants; $\ell=2$, $M=1/2$.}\label{tableIX}
\end{table*}
\begin{table*}
\begin{tabular}{c c c c}
\hline
\hline
$\xi $ & Prony fit & WKB6 Padé & Frobenius  \\
\hline
$0$ & $0.747356-0.177914 i$ & $0.747238-0.177857 i$ & $0.747343-0.177925 i$ \\
$0.2$ & $0.747233-0.178332 i$ & $0.747223-0.178282 i$ & $0.747220-0.178342 i$ \\
$0.4$ & $0.746867-0.179577 i$ & $0.746903-0.179548 i$ & $0.746854-0.179587 i$ \\
$0.6$ & $0.746267-0.181630 i$ & $0.746416-0.181571 i$ & $0.746253-0.181640 i$ \\
$0.8$ & $0.745445-0.184461 i$ & $0.745869-0.184430 i$ & $0.745431-0.184471 i$ \\
$1.$ & $0.744418-0.188031 i$ & $0.745422-0.188359 i$ & $0.744404-0.188042 i$ \\
$1.2$ & $0.743204-0.192298 i$ & $0.744773-0.194024 i$ & $0.743190-0.192309 i$ \\
$1.4$ & $0.741820-0.197217 i$ & $0.742607-0.200808 i$ & $0.741805-0.197228 i$ \\
$1.6$ & $0.740283-0.202745 i$ & $0.739969-0.207172 i$ & $0.740267-0.202756 i$ \\
\hline
\hline
\end{tabular}
\caption{Comparison of the quasinormal frequencies for axial gravitational perturbations ($s=2$) of the second BH model obtained by the time-domain integration and the 6th order WKB approach with Padé approximants: $\ell=2$, $M=1/2$.}\label{tableXI}
\end{table*}
\begin{table*}
\begin{tabular}{c c c c c}
\hline
\hline
$\xi $ &  PS & Frobenius method &  rel. diff. $Re (\omega)$ & rel. diff. $Im (\omega)$\\
\hline
$1$     &    \makecell{ $0.102-3.0901i$} & \makecell{$0.102225-3.090315i$} & \makecell{$0.157\%$}& \makecell{$0.012\%$}\\
\hline
$1.13$&  \makecell{$ 0.1212  -2.5679i $ \\ $ 0.0904  -3.1989i $} & \makecell{ $ 0.121201 - 2.567954i $ \\ $0.08986 - 3.19839i$} & \makecell{$0.0095\%$\\$0.562\%$}& \makecell{$0.0025\%$\\$0.017\%$}\\
\hline
$1.15$ &  \makecell{ $0.05503-1.9585i $ \\ $ 0.124 -2.5799i $} & \makecell{ $0.053611-1.9592i $ \\ $0.124053-2.579967i$} & \makecell{$2.65\%$\\$0.0378\%$}& \makecell{$0.0384\%$\\$0.001\%$}\\
\hline
$1.2$   &  \makecell{ $0.11921-1.973i $ \\ $ 0.126 -2.6106i $} & \makecell{ $ 0.119201 - 1.973059i $ \\ $ 0.125993 - 2.61059i $} & \makecell{$0.0048\%$\\$0.011\%$}& \makecell{$0.001\%$\\$0.0019\%$}\\
\hline
$2$   &  \makecell{ $0.0989-2.6595i $ } & \makecell{ $ 0.099091 - 2.659618i$}& $0.1711\%$ &$0.0037\%$\\
\hline
\hline
\end{tabular}
\caption{Comparison of the nonperturbative quasinormal frequencies for axial gravitational perturbations ($s=2$) of the first BH model obtained by the pseudospectral (PS) method and by the Frobenius method: $\ell=2$, $M=1/2$.}\label{tableXII}
\end{table*}

\section{Quasinormal modes}\label{sec:QNM}

An important preliminary question is determining the range of the quantum correction parameter $\xi$ within which to calculate the quasinormal modes. Since $\xi$ is derived via a perturbative approach, it is not expected to be large. Our criterion for the upper limit is straightforward: if increasing $\xi$ leads to significant changes in the geometry, such that gauge-invariant characteristics, like the fundamental quasinormal mode or the rotational frequency at the innermost stable circular orbit, exhibit more than relatively minor corrections (e.g., several or at most a few tens of percent), then we consider the deviation from the Schwarzschild geometry too strong. At such a point, higher-order corrections should be considered. This criterion guided our choice of the range for $\xi$, which was kept below 1.4 in most cases.

The fundamental quasinormal modes for both black hole models are presented in Tables \ref{tableI}-\ref{tableXI}. A notable difference between the modes of the first and second models is observed in the real oscillation frequency. For the second model, this frequency changes very slightly as the parameter $\xi$ is introduced, whereas it increases with $\xi$ in the first model. This behavior can be attributed to the differences in the black hole metrics: in the first model, both the components $g_{tt}$ and $g_{rr}$ differ from the Schwarzschild case, while in the second model, $g_{tt}$ retains the Schwarzschild form. As a result, the real part of the frequency, $\text{Re}(\omega)$, which is primarily determined by the centrifugal part of the effective potential, exhibits only minor changes with $\xi$.

It can be observed that the real oscillation frequency of the fundamental mode increases monotonically with the parameter $\xi$ for perturbations of all types of the first model and for scalar perturbations of the second model. However,  this is not the case of the electromagnetic and gravitational perturbations of the second black hole model. Such behavior could be explained by the form of the effective potentials which are higher for larger values of $\xi$ (see figs. \ref{fig:potentials1}-\ref{fig:potentials4} ), while this is not so for the  perturbations of the second model, as can be seen directly from the analytic expression for the potential 
fig. \ref{fig:potentials5}. 
The damping rate also generally increases with $\xi$, up to a certain relatively large value, beyond which it may start to decrease, as illustrated in Fig. \ref{fig:BH1L1s1}. A comparison of the WKB approach, time-domain integration, and the precise results of the Frobenius method indicates that while the WKB method can be more accurate for $\ell=0$, the time-domain integration becomes more reliable for larger values of $\ell$. This difference in accuracy arises because, at $\ell=0$, the ringing period is very short, making it challenging for the Prony method to extract frequencies with high precision.

The most striking effect is not observed in the fundamental mode, which changes relatively smoothly with the parameter $\xi$, but rather in the first few overtones, which deviate from their Schwarzschild values at an increasing rate. As depicted in the figures, while the fundamental mode only changes by a few percent, the first overtone can change by orders of magnitude. Moreover, we observe a qualitative change in the spectrum of overtones, as the real part of the frequency, $\text{Re}(\omega)$, tends to zero. A similar phenomenon of the vanishing real oscillation frequency of the overtones in the presence of quantum corrections was recently observed in \cite{Zinhailo:2024kbq}. This vanishing real part of the frequency can also occur for massive fields \cite{Zinhailo:2024jzt,Konoplya:2005hr}.

The observed high sensitivity of the overtones compared to the fundamental mode can be explained as follows: while the fundamental mode is primarily determined by the behavior of the effective potential near the peak of the potential barrier and is relatively insensitive to near-horizon deformations, the overtones are highly sensitive even to slight near-horizon deformations \cite{Konoplya:2022pbc,Konoplya:2023hqb}. This phenomenon, which can be thought of as the "sound" of the event horizon, has been termed the "outburst of overtones" and has recently been the subject of several studies \cite{Konoplya:2022hll,Konoplya:2022iyn,Konoplya:2023aph,Konoplya:2023ppx,Bolokhov:2023bwm,Zhang:2024nny}.
It is interesting to note that, with respect to the behavior of overtones, the two models also exhibit differences: the deviations of overtones from their Schwarzschild values for gravitational perturbations develop much more slowly in the second black hole model than in the first one.

In addition to the quasinormal modes described above, which transition into the Schwarzschild modes as the quantum parameter approaches zero, we observe some frequencies (see, for example, fig. \ref{fig:Non-Perturbative}) that do not transition to the Schwarzschild frequencies. These modes may become purely imaginary, i.e., non-oscillatory, at certain values of $\xi$. Being non-perturbative in $\xi$, these modes cannot definitively be attributed to the quantum-corrected black hole, as the metric for the latter is constructed perturbatively in $\xi$.

Following the general procedure outlined in \cite{Konoplya:2023moy}, we can derive an exact analytical formula for the quasinormal frequencies in the eikonal limit, $\ell \gg n$. By expanding in powers of the inverse quantity $\kappa = \ell + 1/2$, we locate the position of the maximum of the effective potential. Remarkably, for both the first and second black hole models, this position coincides with that of the Schwarzschild case:
\begin{equation}
r_{max} = 3 M + O\left(\frac{1}{\kappa}\right).
\end{equation}
Then, using the first order WKB formula and the above expression for $r_{max}$ we can find the frequencies in the form of expansion in terms of powers of $\xi$;
\begin{widetext}
\begin{equation}\label{eikonal1}
\omega_{n} =\frac{\kappa}{3 \sqrt{3}
   M}-\frac{i (2 n+1)}{6
   \sqrt{3}
   M}+\xi ^2
   \left(\frac{\kappa}{162 \sqrt{3}
   M^3}-\frac{i (2 n+1)}{108
   \sqrt{3}
   M^3}\right)+\xi ^4
   \left(-\frac{\kappa}{17496
   \sqrt{3}
   M^5}+\frac{i (2  n+1)}{34992 \sqrt{3} M^5}\right)+O\left(\xi ^5, \frac{1}{\kappa}\right).
   \end{equation}
In a similar way for the second black hole model, we have
\begin{equation}\label{eikonal2}
\omega_{n} =\frac{\kappa}{3 \sqrt{3}
   M}-\frac{i (2 n+1)}{6 \sqrt{3}M}-\xi ^2
   \frac{i (2 n+1)}{324 \sqrt{3} M^3}+\xi ^4
   \frac{i (2 n+1)}{34992 \sqrt{3} M^5}+O\left(\xi ^5, \frac{1}{\kappa}\right).
\end{equation}
\end{widetext}
When $\xi =0$, the above formulas reduce to the well-known expressions for the Shwarzschild black hole \cite{Blome:1981azp}.

It is worth mentioning that the above eikonal expressions for quasinormal modes are related to the parameters of the unstable null geodesics, such as the rotational frequency and Lyapunov exponent, via the correspondence established in \cite{Cardoso:2008bp}. However, there are a number of counterexamples to this correspondence \cite{Konoplya:2019hml,Konoplya:2020bxa,Bolokhov:2023dxq}. In \cite{Konoplya:2017wot,Konoplya:2022gjp} it was shown that the correspondence works only when the WKB approach can be applied and only to the part of the eikonal spectrum which can be found by the WKB method. 
Here we see that the correspondence takes place for both black hole models.

Within the recently discussed link between grey-body factors and amplitude of gravitational waves \cite{Oshita:2023cjz}, a correspondence between the grey-body factors and quasinormal modes has been established \cite{Konoplya:2024lir}:
\begin{equation}\label{transmission-eikonal}
\Gamma_{\ell}(\omega)=\left(1+e^{2\pi\dfrac{\omega^2-\re{\omega_0}^2}{4\re{\omega_0}\im{\omega_0}}}\right)^{-1} + \Sigma(\omega_{0},\omega_{1}).\\
\end{equation}
Here $\Sigma(\omega_{0},\omega_{1})$ is the sum of the correction terms beyond the eikonal limit found in \cite{Konoplya:2024lir}.

\hfill \break
This relation, which connects the grey-body factors $\Gamma_{\ell}(\omega)$ with the fundamental quasinormal mode $\omega_0$, is exact in the eikonal limit as $\ell \to \infty$. By introducing corrections involving the first overtone $\omega_1$, this relation also provides an approximate correspondence for lower $\ell$. Consequently, the quasinormal modes obtained here can be used to determine the grey-body factors. In the eikonal regime, the grey-body factors can be immediately derived analytically using the eikonal expressions (\ref{eikonal1}) and (\ref{eikonal2}) for the quasinormal modes $\omega_1$ and $\omega_2$ in (\ref{transmission-eikonal}). This correspondence also applies to a variety of rotating black holes \cite{Konoplya:2024vuj}.

It is worth mentioning that, at asymptotically late times, the quasinormal ringing transitions to power-law tails, which are indistinguishable from the Price law for the Schwarzschild solution \cite{Price:1971fb} (see fig. \ref{fig:TD1}),
\begin{equation}
|\Psi| \sim t^{-(2 \ell+3)}, \quad t \rightarrow \infty.
\end{equation}
\section{Testing agnostic parametrization}
Recently, several parameterized agnostic frameworks to probe deviations in the QNM spectra in modified gravity have been developed \cite{Cardoso_2019, Franchini_2023, McManus_2019, cano2024parametrizedquasinormalmodeframework}. Following \cite{Cardoso_2019}, the quasinormal frequency can be written as 
\eq{
\omega=\omega_{\rm 0} + \sum_{j=0}^{\infty} \alpha_j e_j,
}
where $\omega_{\rm 0}$\footnote{In the units of the BH radius} is the corresponding frequency of Schwarzchild black hole, $e_j$ some universal coefficients and $\alpha_j$ are coefficient of expansion of effective potential up to first order, i.e
\eq{V=V_{GR}+\delta V,\, \delta V=   \frac{1}{r_h^2} \sum_{j=0}^{\infty} \alpha_j \left( \frac{r_h}{r} \right)^j,
}
The considered in this paper metrics are relatively simple and are within a class that can be described by these approach. The goal of this section is to test  accuracy of this approach to describe the fundamental mode and the first overtones of the perturbative branches.

Following \cite{Cardoso_2019}, we can assume that
\eq{\label{eq:param_metric}
f(r)=f_{(0)}(r)(1+\sigma_1(r)),\,\,g(r)=f_{(0)}(r)(1+\sigma_2(r)),}
where $\sigma_1(r)$ and $\sigma_2(r)$ are small corrections and $f_{(0)}(r)=1-r_h/r$.

It is possible to rewrite the master equation (\ref{wave-equation}) in the form
\eq{
f_{(0)}\frac{d}{dr} \left[ f_{(0)} \frac{d\phi}{dr} \right] + \left[ \frac{\omega^2}{1+2\delta Z} - f_{0} (\tilde{V}+\delta\tilde{V})\right] \phi = 0, 
}
with 
\eq{
\tilde{V} = \bar{V}_{GR} + \delta \bar{V} - \bar{V}_{GR} \delta Z + \frac{1}{2}(f_{(0)} \delta Z')',
}
where $\phi=\sqrt{1+\delta Z}\Psi$, $\bar{V}=V/{f_{(0)}(1+\delta Z)}$, and $\delta Z=(\sigma_1+\sigma_2)/2$.

For the frequency-dependent term, we have

\eq{
\frac{\omega^2}{1+2\delta Z} = \omega^2 \left[ 1 - 2 \delta Z(r_h) \right] - 2 \omega^2 \left[ \delta Z(r) - \delta Z(r_h) \right].
}
Therefore, the perturbed part  $\delta\tilde{V}$ of  effective potential has the following form
\eq{
\delta\tilde{V}=\delta \bar{V} - \bar{V}_{GR} \delta Z + \frac{1}{2}(f_{(0)} \delta Z')'+ \frac{2 \omega_0^2}{f_{(0)}} \left[ \delta Z(r) - \delta Z(r_H)\right].
}
Then, the resulting quasinormal frequency can be written as
\eq{
\omega_{\rm } = \left( 1 + \delta Z(r_h) \right)\left[\omega_{\rm 0} + \sum_{j=0}^{\infty} \alpha_j e_j\right],
}
In our case for the BH model 1, we have
\eq{
\sigma_1(r)=\sigma_2(r)=\left(1-\frac{2M}{r}\right)\frac{\xi ^2}{r^2}.
}
and for the BH model 2, we have
\eq{
\sigma_1(r)=0,\quad \sigma_2(r)=\left(1-\frac{2M}{r}\right)\frac{\xi ^2}{r^2},
}

In addition to the two black hole models in effective quantum gravity that we study here, we will also test the applicability of the agnostic parametrization by calculating quasinormal modes for axial gravitational perturbations of the black hole obtained within the Quantum Oppenheimer-Snyder model~\cite{Lewandowski_2023}. While quasinormal modes of test fields have been recently considered in a few works for this model~\cite{Gong:2023ghh,Zinhailo:2024kbq,Luo:2024dxl}, gravitational perturbations have not been considered so far. In this way, we complement the existing literature on quasinormal modes of such a quantum-corrected black hole model by adding the most important gravitational sector.

The metric has the following form model~\cite{Lewandowski_2023},
\eq{\label{eq:metric_3}
f(r)=g(r)=1-\frac{2M}{r}+\frac{M^2\xi}{r^4},
}
where $\xi$ is the Barbero-Immirzi parameter and we have a black hole solution for $\xi\leq 27M^2/16$, respectively. For simplicity we put $M=1$ for this case.
Similarly to the previous cases, we can transform metric (\ref{eq:metric_3}) into the form (\ref{eq:param_metric}) up to the first order in $\xi$ with  
\eq{
\sigma_1(r)= \sigma_2(r)=-  \left(\frac{1}{8 r}+\frac{1}{4 r^2}+\frac{1}{2 r^3}\right)\xi,
}

\begin{figure*}
\resizebox{\linewidth}{!}{\includegraphics{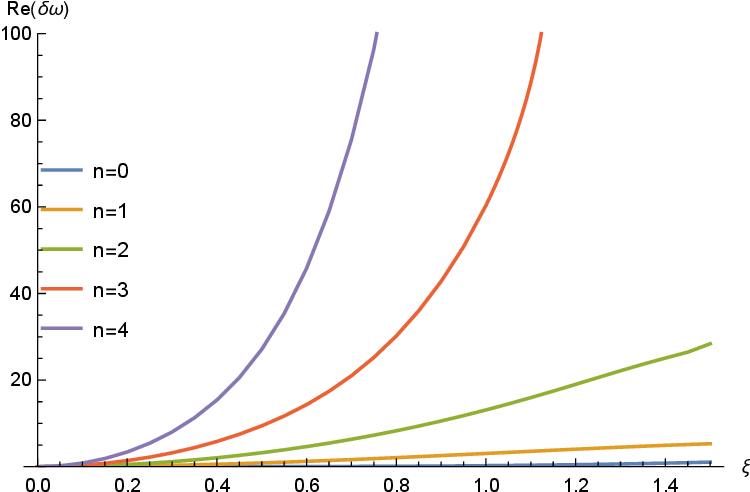}~~~\includegraphics{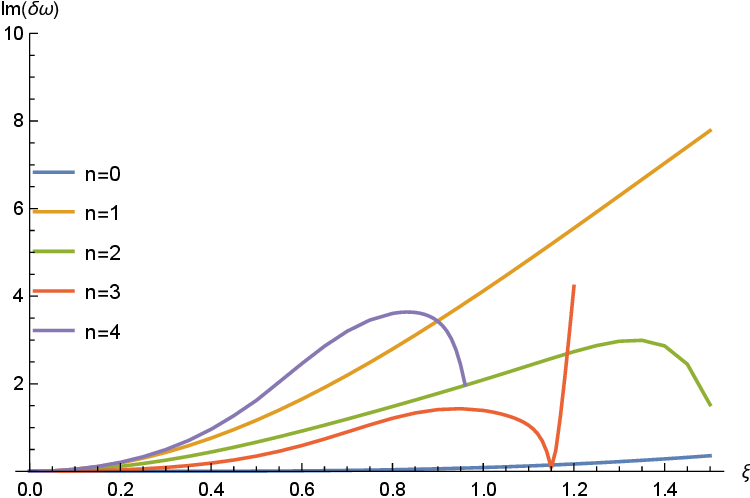}}
\caption{The relative differences between the exact QNM values and those obtained via the agnostic parameterization for axial perturbations in model 1; $\ell=2$, $M=1/2$.}\label{fig:prm1}
\end{figure*}

The results of the comparison between the exact quasinormal modes, obtained via the Frobenius method, and the corresponding agnostic parameterization for axial gravitational perturbations are shown in Figs. \ref{fig:prm1}--\ref{fig:prm2} and in Tabs. \ref{tablecomp1}--\ref{tablecomp2}. 
Comparison of the precise quasinormal modes with those found via the parametrization shows that the latter could be used for reasonable estimation of the fundamental mode for the first and second models in the effective quantum gravity at moderate values of the coupling, because in those cases the relative error is one or more orders smaller than the effect, that is, the deviation of the frequencies from their Schwarzschild limits.  However, for the quantum Oppenheimer-Snyder model developed in \cite{Lewandowski:2022zce} such an accuracy is achieved only in the regime of very small coupling constant, say $\xi \sim 0.1$, while for larger $\xi$ the relative error quickly becomes of the same order as an effect.

However, for the  overtones, this parameterization fails already at the first overtone and small coupling constant $\xi \sim 0.1$, producing the error of the same order or higher than the effect, while the frequencies for such small couplings are practically indistinguishable from  the Schwarzschild ones.
This failure of the parametrization at overtones is expected because of the high sensitivity of overtones to the near horizon deformations and spectral instability of QNMs, leading to less reliable results for those modes. The similar results about poor performance of this agnostic approach to describe overtones  for another spacetimes were obtained in \cite{2024PhRvD.110b4015H}.  Additionally, it should be noted that this approach is in principle inapplicable to non-perturbative branches, due to the absence of a corresponding $\omega_0$ in GR case.

Although the accuracy of the parametrization can be improved by introducing non-linear terms \cite{McManus_2019} or by using Padé approximants \cite{2024PhRvD.110b4015H}, this approach effectively replaces the 
complex yet direct  and precise calculations
with other complicated methods that still introduce some error and cannot describe QNMs spectra with sufficient accuracy. Moreover, then the original simplicity of the method is lost and the initial idea of effectively constraining the geometry becomes impractical. As a method for calculating quasinormal modes of some classes of metrics, we also find it ineffective because the accuracy depends on the black hole model under consideration and, as we showed here, may be insufficient even for small deviations from the Schwarzschild limit. 
\begin{figure*}
\resizebox{\linewidth}{!}{\includegraphics{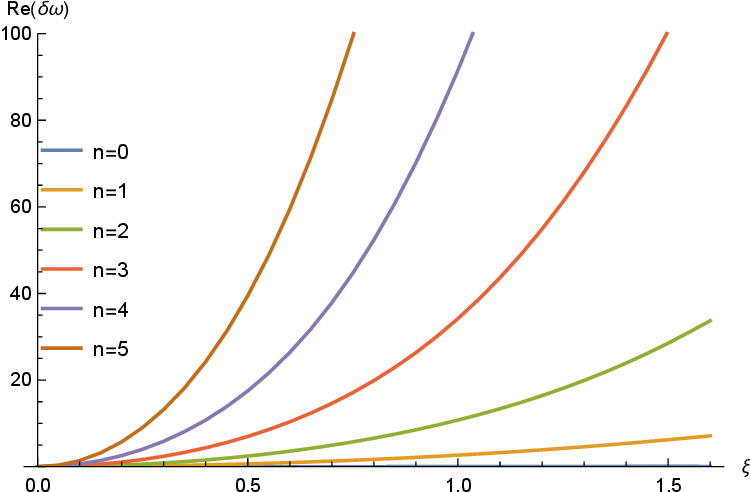}~~~\includegraphics{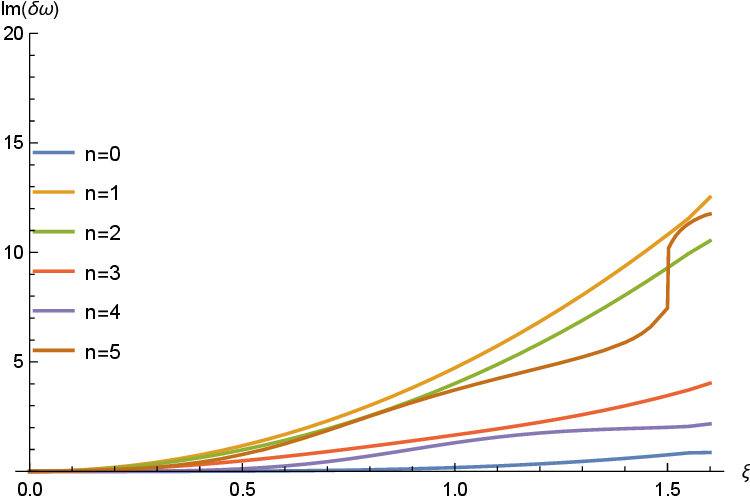}}
\caption{The relative differences between the exact QNM values and those obtained via the agnostic parameterization for axial perturbations in model 1; $\ell=2$, $M=1/2$.}\label{fig:prm2}
\end{figure*}
\begin{table*}
\begin{tabular}{c c c c c c c}
\hline
\hline
$\xi $ &  Frobenius method & Parametrization  &  $\delta Re (\omega)$ error & $ \delta Im (\omega)$ error & $\delta  Re(\omega)$ effect &
$\delta Im (\omega)$ effect \\
\hline
$0.1$ & \makecell{$0.747787-0.178260 i$ \\ $0.693426-0.548835 i$ \\ \
$0.601199-0.958268 i$ \\ $0.50077-1.41286 i$ \\ $0.41120-1.89731 i$} \
& \makecell{$0.747787-0.178260 i$ \\ $0.693173-0.549103 i$ \\ \
$0.600422-0.958550 i$ \\ $0.49903-1.41275 i$ \\ $0.40779-1.89636 i$} \
& \makecell{$0.00002659$ \\ $0.03645$ \\ $0.1292$ \\ $0.3475$ \\ \
$0.8275$} & \makecell{$0.00001075$ \\ $0.04889$ \\ $0.02940$ \\ \
$0.008026$ \\ $0.04995$} & \makecell{$0.05939$ \\ $0.0005642$ \\ \
$0.1508$ \\ $0.4463$ \\ $0.9234$} & \makecell{$0.1887$ \\ $0.1835$ \\ \
$0.1792$ \\ $0.1821$ \\ $0.1910$} \\
\hline
$0.2$ & \makecell{$0.749116-0.179267 i$ \\ $0.693433-0.551850 i$ \\ \
$0.598460-0.963418 i$ \\ $0.49399-1.42061 i$ \\ $0.39966-1.90833 i$} \
& \makecell{$0.749119-0.179267 i$ \\ $0.692427-0.552923 i$ \\ \
$0.595368-0.964537 i$ \\ $0.48707-1.42011 i$ \\ $0.38609-1.90437 i$} \
& \makecell{$0.0004239$ \\ $0.1451$ \\ $0.5166$ \\ $1.401$ \\ \
$3.396$} & \makecell{$0.0001713$ \\ $0.1944$ \\ $0.1162$ \\ $0.03509$ \
\\ $0.2076$} & \makecell{$0.2372$ \\ $0.001517$ \\ $0.6057$ \\ \
$1.793$ \\ $3.702$} & \makecell{$0.7546$ \\ $0.7339$ \\ $0.7176$ \\ \
$0.7315$ \\ $0.7730$} \\
\hline
$0.5$ & \makecell{$0.758321-0.186306 i$ \\ $0.693264-0.572970 i$ \\ \
$0.578629-0.999774 i$ \\ $0.44542-1.47683 i$ \\ $0.32121-1.99276 i$} \
& \makecell{$0.758444-0.186318 i$ \\ $0.687201-0.579664 i$ \\ \
$0.559988-1.006449 i$ \\ $0.40339-1.47166 i$ \\ $0.23415-1.96042 i$} \
& \makecell{$0.01614$ \\ $0.8746$ \\ $3.221$ \\ $9.437$ \\ $27.10$} & \
\makecell{$0.006417$ \\ $1.168$ \\ $0.6677$ \\ $0.3504$ \\ $1.623$} & \
\makecell{$1.469$ \\ $0.02273$ \\ $3.899$ \\ $11.45$ \\ $22.61$} & \
\makecell{$4.710$ \\ $4.589$ \\ $4.518$ \\ $4.718$ \\ $5.232$} \\
\hline
$0.8$ & \makecell{$0.774977-0.199333 i$ \\ $0.691964-0.612222 i$ \\ \
$0.538979-1.068452 i$ \\ $0.35518-1.58663 i$ \\ $0.19972-2.14176 i$} \
& \makecell{$0.775760-0.199411 i$ \\ $0.677497-0.629325 i$ \\ \
$0.494283-1.084286 i$ \\ $0.24798-1.56738 i$ \\ $-0.04801-2.06452 i$} \
& \makecell{$0.1010$ \\ $2.091$ \\ $8.293$ \\ $30.18$ \\ $124.0$} & \
\makecell{$0.03894$ \\ $2.793$ \\ $1.482$ \\ $1.213$ \\ $3.607$} & \
\makecell{$3.698$ \\ $0.2103$ \\ $10.48$ \\ $29.39$ \\ $51.88$} & \
\makecell{$12.03$ \\ $11.75$ \\ $11.70$ \\ $12.50$ \\ $13.10$} \\
\hline
$1.$ & \makecell{$0.789874-0.211309 i$ \\ $0.689629-0.648458 i$ \\ \
$0.499042-1.132457 i$ \\ $0.26359-1.67911 i$ \\ $0.17483-2.16061 i$} \
& \makecell{$0.791744-0.211496 i$ \\ $0.668539-0.675166 i$ \\ \
$0.433632-1.156135 i$ \\ $0.10453-1.65574 i$ \\ $-0.09547-2.08203 i$} \
& \makecell{$0.2368$ \\ $3.058$ \\ $13.11$ \\ $60.35$ \\ $154.6$} & \
\makecell{$0.08871$ \\ $4.119$ \\ $2.091$ \\ $1.392$ \\ $3.637$} & \
\makecell{$5.691$ \\ $0.5469$ \\ $17.12$ \\ $47.60$ \\ $57.87$} & \
\makecell{$18.76$ \\ $18.37$ \\ $18.39$ \\ $19.06$ \\ $14.10$} \\
\hline

\end{tabular}
\caption{Comparison of the quasinormal frequencies for axial gravitational perturbations ($s=2$) of the first BH model obtained using the Frobenius  method and  the agnostic parametrization: $\ell=2$, $M=1/2$. The columns "error" stand for the relative difference (in percents) between precise values of the QNMs and those obtained via the parametrization, while the "effect" columns designate the relative difference in percents between the values of QNMs for the Schwarzschild and quantum corrected black holes. }\label{tablecomp1}
\end{table*}

\begin{table*}
\begin{tabular}{c c c c c c c}
\hline
\hline
$\xi $ &  Frobenius method & Parametrization  &  $\delta Re (\omega)$ error & $ \delta Im (\omega)$ error & $\delta  Re(\omega)$ effect &
$\delta Im (\omega)$ effect  \\
\hline
$0.1$ & \makecell{$0.747313-0.178029 i$ \\ $0.693226-0.548147 i$ \\ $0.601587-0.957116 i$ \\ $0.50208-1.41118 i$ \\ $0.41377-1.89496 i$ \\ $0.33737-2.39287 i$} & \makecell{$0.747313-0.178029 i$ \\ $0.693045-0.548403 i$ \\ $0.601012-0.957488 i$ \\ $0.50076-1.41146 i$ \\ $0.41116-1.89496 i$ \\ $0.33261-2.39248 i$} & \makecell{$2\cdot10^{-6}$ \\ $0.026$ \\ $0.096$ \\ $0.26$ \\ $0.63$ \\ $1.4$} & \makecell{$0.000020$ \\ $0.047$ \\ $0.039$ \\ $0.020$ \\ $0.000079$ \\ $0.016$} & \makecell{$0.004122$ \\ $0.02833$ \\ $0.08639$ \\ $0.1847$ \\ $0.3041$ \\ $0.3632$} & \makecell{$0.05869$ \\ $0.05790$ \\ $0.05873$ \\ $0.06244$ \\ $0.06705$ \\ $0.06935$} \\
\hline
$0.2$ & \makecell{$0.747220-0.178342 i$ \\ $0.692636-0.549096 i$ \\ $0.600028-0.958797 i$ \\ $0.49931-1.41382 i$ \\ $0.41002-1.89881 i$ \\ $0.33375-2.39805 i$} & \makecell{$0.747220-0.178343 i$ \\ $0.691913-0.550123 i$ \\ $0.597726-0.960292 i$ \\ $0.49403-1.41495 i$ \\ $0.39955-1.89877 i$ \\ $0.31465-2.39627 i$} & \makecell{$0.000034$ \\ $0.10$ \\ $0.38$ \\ $1.1$ \\ $2.6$ \\ $5.7$} & \makecell{$0.00032$ \\ $0.19$ \\ $0.16$ \\ $0.080$ \\ $0.0021$ \\ $0.074$} & \makecell{$0.01646$ \\ $0.1133$ \\ $0.3453$ \\ $0.7364$ \\ $1.206$ \\ $1.431$} & \makecell{$0.2345$ \\ $0.2312$ \\ $0.2345$ \\ $0.2500$ \\ $0.2706$ \\ $0.2857$} \\
\hline
$0.5$ & \makecell{$0.746582-0.180514 i$ \\ $0.688506-0.555654 i$ \\ $0.589197-0.970394 i$ \\ $0.48044-1.43247 i$ \\ $0.38572-1.92776 i$ \\ $0.31221-2.44173 i$} & \makecell{$0.746573-0.180536 i$ \\ $0.683991-0.562162 i$ \\ $0.574727-0.979917 i$ \\ $0.44688-1.43939 i$ \\ $0.31827-1.92547 i$ \\ $0.18892-2.42282 i$} & \makecell{$0.0013$ \\ $0.66$ \\ $2.5$ \\ $7.0$ \\ $17.$ \\ $39.$} & \makecell{$0.012$ \\ $1.2$ \\ $0.98$ \\ $0.48$ \\ $0.12$ \\ $0.77$} & \makecell{$0.1018$ \\ $0.7090$ \\ $2.144$ \\ $4.487$ \\ $7.063$ \\ $7.793$} & \makecell{$1.455$ \\ $1.428$ \\ $1.447$ \\ $1.572$ \\ $1.799$ \\ $2.112$} \\
\hline
$0.8$ & \makecell{$0.745431-0.184471 i$ \\ $0.680816-0.567428 i$ \\ $0.569531-0.991162 i$ \\ $0.44848-1.46798 i$ \\ $0.35115-1.98910 i$ \\ $0.29131-2.53608 i$} & \makecell{$0.745371-0.184611 i$ \\ $0.669278-0.584520 i$ \\ $0.532014-1.016363 i$ \\ $0.35931-1.48478 i$ \\ $0.16733-1.97505 i$ \\ $-0.04458-2.47212 i$} & \makecell{$0.0081$ \\ $1.7$ \\ $6.6$ \\ $20.$ \\ $52.$ \\ $120.$} & \makecell{$0.076$ \\ $3.0$ \\ $2.5$ \\ $1.1$ \\ $0.71$ \\ $2.5$} & \makecell{$0.2558$ \\ $1.818$ \\ $5.410$ \\ $10.84$ \\ $15.39$ \\ $13.97$} & \makecell{$3.679$ \\ $3.578$ \\ $3.618$ \\ $4.090$ \\ $5.038$ \\ $6.058$} \\
\hline
$1.$ & \makecell{$0.744404-0.188042 i$ \\ $0.673700-0.577846 i$ \\ $0.552025-1.009493 i$ \\ $0.42321-1.50172 i$ \\ $0.33120-2.04764 i$ \\ $0.28042-2.61494 i$} & \makecell{$0.744261-0.188372 i$ \\ $0.655698-0.605159 i$ \\ $0.492587-1.050005 i$ \\ $0.27849-1.52668 i$ \\ $0.02800-2.02081 i$ \\ $0.26012-2.51764 i$} & \makecell{$0.019$ \\ $2.7$ \\ $11.$ \\ $34.$ \\ $92.$ \\ $190.$} & \makecell{$0.18$ \\ $4.7$ \\ $4.0$ \\ $1.7$ \\ $1.3$ \\ $3.7$} & \makecell{$0.3933$ \\ $2.844$ \\ $8.318$ \\ $15.86$ \\ $20.20$ \\ $17.18$} & \makecell{$5.686$ \\ $5.479$ \\ $5.534$ \\ $6.483$ \\ $8.129$ \\ $9.356$} \\
\hline

\end{tabular}
\caption{Comparison of the quasinormal frequencies for axial gravitational perturbations ($s=2$) of the second BH model obtained using the Frobenius  method and  the agnostic parametrization: $\ell=2$, $M=1/2$. The columns "error" stand for the relative difference (in percents) between precise values of the QNMs and those obtained via the parametrization, while the "effect" columns designate the relative difference in percents between the values of QNMs for the Schwarzschild and quantum corrected black holes.}\label{tablecomp2}
\end{table*}

\begin{figure*}
\resizebox{\linewidth}{!}{\includegraphics{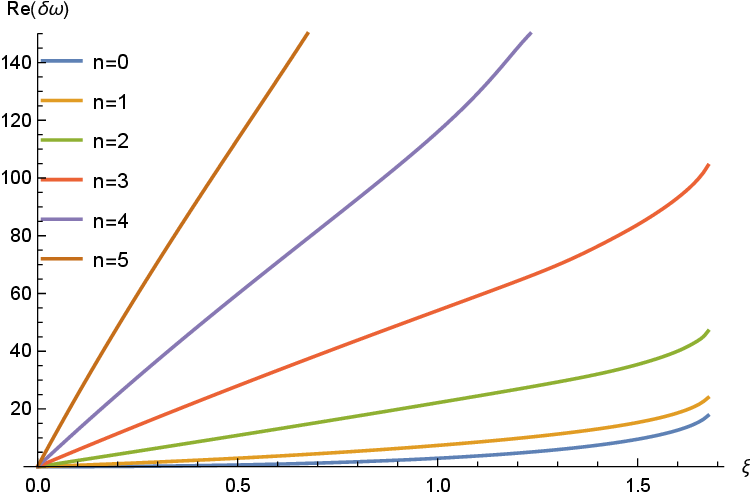}~~~\includegraphics{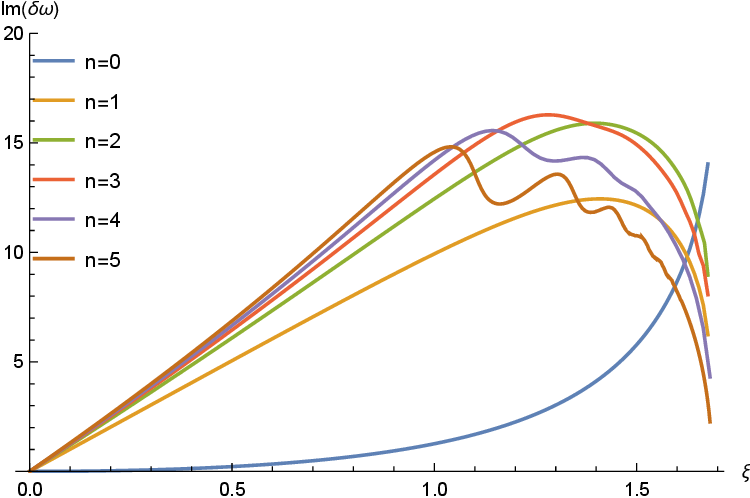}}
\caption{The relative differences between the exact QNM values and those obtained via the agnostic parameterization for axial perturbations in third model; $\ell=2$, $M=1$.}\label{fig:prm2}
\end{figure*}
\begin{figure*}
\resizebox{\linewidth}{!}{\includegraphics{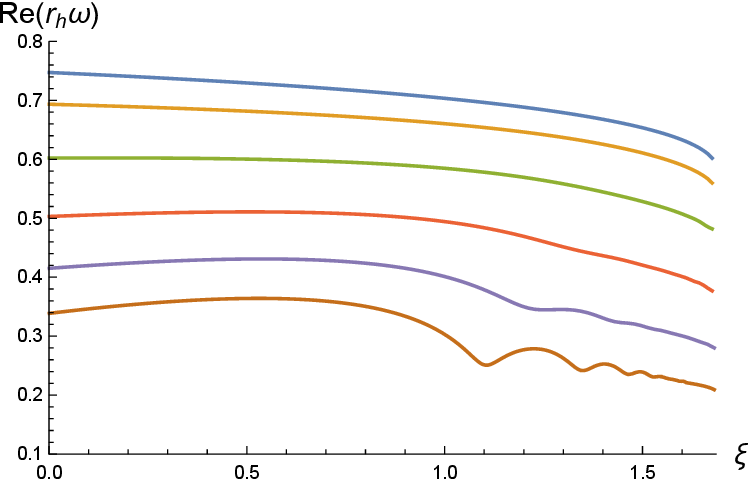}~~~\includegraphics{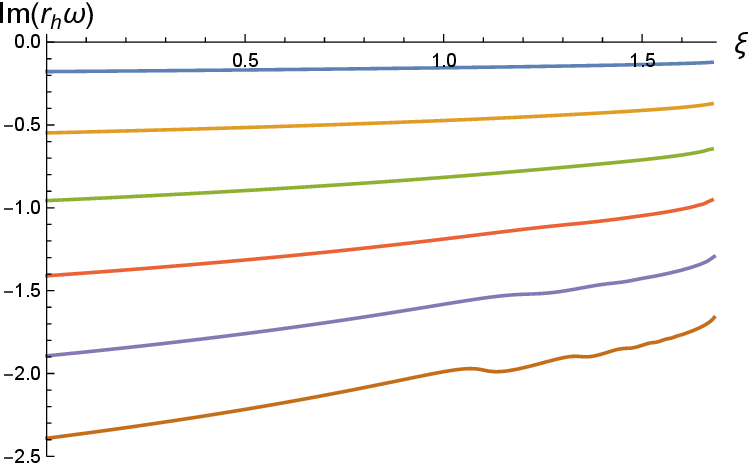}}
\caption{The fundamental mode and the first five overtones as a function of $\xi$ for the third black hole model; $\ell=2$, $M=1$ axial gravitational perturbations.}\label{fig:NNBH2L2s2}
\end{figure*}
\begin{table*}
\begin{tabular}{c c c c c c c}
\hline
\hline
$\xi $ &  Frobenius method & Parametrization  &  $\delta Re (\omega)$ error & $ \delta Im (\omega)$ error & $\delta  Re(\omega)$ effect &
$\delta Im (\omega)$ effect  \\
\hline
$0.1$ & \makecell{$0.374489-0.088596 i$ \\ $0.347981-0.272687 i$ \\ $0.303097-0.475797 i$ \\ $0.254362-0.700981 i$ \\ $0.211209-0.940775 i$ \\ $0.174244-1.187497 i$} & \makecell{$0.374409-0.088589 i$ \\ $0.346245-0.275447 i$ \\ $0.296617-0.481486 i$ \\ $0.239640-0.709718 i$ \\ $0.184345-0.952727 i$ \\ $0.130838-1.202944 i$} & \makecell{$0.02133$ \\ $0.4989$ \\ $2.138$ \\ $5.788$ \\ $12.72$ \\ $24.91$} & \makecell{$0.007259$ \\ $1.012$ \\ $1.196$ \\ $1.246$ \\ $1.270$ \\ $1.301$} & \makecell{$0.4216$ \\ $0.2739$ \\ $0.03750$ \\ $0.4892$ \\ $1.119$ \\ $2.215$} & \makecell{$1.057$ \\ $1.094$ \\ $1.166$ \\ $1.239$ \\ $1.290$ \\ $1.329$} \\
\hline
$0.3$ & \makecell{$0.376166-0.087798 i$ \\ $0.350545-0.269995 i$ \\ $0.307128-0.470318 i$ \\ $0.259802-0.691723 i$ \\ $0.217919-0.927222 i$ \\ $0.182757-1.169364 i$} & \makecell{$0.375403-0.087735 i$ \\ $0.344846-0.278193 i$ \\ $0.287292-0.487343 i$ \\ $0.215444-0.718019 i$ \\ $0.137484-0.963348 i$ \\ $0.053293-1.216156 i$} & \makecell{$0.2028$ \\ $1.626$ \\ $6.458$ \\ $17.07$ \\ $36.91$ \\ $70.84$} & \makecell{$0.07144$ \\ $3.036$ \\ $3.620$ \\ $3.801$ \\ $3.896$ \\ $4.002$} & \makecell{$1.356$ \\ $0.9163$ \\ $0.01415$ \\ $1.226$ \\ $2.839$ \\ $5.481$} & \makecell{$3.385$ \\ $3.513$ \\ $3.759$ \\ $4.012$ \\ $4.191$ \\ $4.322$} \\
\hline
$0.5$ & \makecell{$0.377901-0.086899 i$ \\ $0.353112-0.266936 i$ \\ $0.310976-0.464030 i$ \\ $0.264612-0.681010 i$ \\ $0.223188-0.911432 i$ \\ $0.188540-1.148191 i$} & \makecell{$0.375637-0.086708 i$ \\ $0.342711-0.280430 i$ \\ $0.277249-0.492308 i$ \\ $0.190509-0.724987 i$ \\ $0.089794-0.972156 i$ \\ $0.025238-1.227053 i$} & \makecell{$0.5992$ \\ $2.946$ \\ $10.85$ \\ $28.00$ \\ $59.77$ \\ $86.61$} & \makecell{$0.2202$ \\ $5.055$ \\ $6.094$ \\ $6.458$ \\ $6.663$ \\ $6.868$} & \makecell{$2.440$ \\ $1.722$ \\ $0.2942$ \\ $1.532$ \\ $3.675$ \\ $6.973$} & \makecell{$6.060$ \\ $6.309$ \\ $6.781$ \\ $7.272$ \\ $7.625$ \\ $7.878$} \\
\hline
$0.7$ & \makecell{$0.379692-0.085880 i$ \\ $0.355628-0.263434 i$ \\ $0.314431-0.456763 i$ \\ $0.268237-0.668532 i$ \\ $0.225830-0.892922 i$ \\ $0.189415-1.123298 i$} & \makecell{$0.374910-0.085461 i$ \\ $0.339645-0.282024 i$ \\ $0.266302-0.496145 i$ \\ $0.164641-0.730268 i$ \\ $0.041058-0.978670 i$ \\ $0.105017-1.235020 i$} & \makecell{$1.260$ \\ $4.494$ \\ $15.31$ \\ $38.62$ \\ $81.82$ \\ $44.56$} & \makecell{$0.4881$ \\ $7.057$ \\ $8.622$ \\ $9.235$ \\ $9.603$ \\ $9.946$} & \makecell{$3.728$ \\ $2.756$ \\ $0.9151$ \\ $1.176$ \\ $3.149$ \\ $5.794$} & \makecell{$9.182$ \\ $9.592$ \\ $10.36$ \\ $11.17$ \\ $11.76$ \\ $12.18$} \\
\hline
$1$ & \makecell{$0.382466-0.084069 i$ \\ $0.359062-0.257165 i$ \\ $0.317938-0.443685 i$ \\ $0.268767-0.646126 i$ \\ $0.218098-0.860190 i$ \\ $0.164787-1.081552 i$} & \makecell{$0.371284-0.083007 i$ \\ $0.332595-0.282710 i$ \\ $0.247501-0.498910 i$ \\ $0.123387-0.733728 i$ \\ $0.034794-0.982378 i$ \\ $0.227963-1.239237 i$} & \makecell{$2.924$ \\ $7.371$ \\ $22.15$ \\ $54.09$ \\ $84.05$ \\ $38.34$} & \makecell{$1.264$ \\ $9.933$ \\ $12.45$ \\ $13.56$ \\ $14.20$ \\ $14.58$} & \makecell{$6.237$ \\ $4.998$ \\ $2.963$ \\ $1.754$ \\ $3.461$ \\ $11.72$} & \makecell{$15.07$ \\ $15.82$ \\ $17.22$ \\ $18.67$ \\ $19.69$ \\ $20.20$} \\
\hline
$1.5$ & \makecell{$0.387094-0.080057 i$ \\ $0.361786-0.244024 i$ \\ $0.312990-0.420692 i$ \\ $0.249079-0.620366 i$ \\ $0.187914-0.844646 i$ \\ $0.141578-1.083589 i$} & \makecell{$0.350234-0.075425 i$ \\ $0.306372-0.273659 i$ \\ $0.202164-0.485686 i$ \\ $0.040268-0.712917 i$ \\ $0.177262-0.952494 i$ \\ $0.451974-1.200325 i$} & \makecell{$9.522$ \\ $15.32$ \\ $35.41$ \\ $83.83$ \\ $5.668$ \\ $219.2$} & \makecell{$5.785$ \\ $12.14$ \\ $15.45$ \\ $14.92$ \\ $12.77$ \\ $10.77$} & \makecell{$14.36$ \\ $13.53$ \\ $13.95$ \\ $19.62$ \\ $30.82$ \\ $41.66$} & \makecell{$31.64$ \\ $32.97$ \\ $34.68$ \\ $34.65$ \\ $32.80$ \\ $30.71$} \\
\hline

\end{tabular}
\caption{Comparison of the quasinormal frequencies for axial gravitational perturbations ($s=2$) of the third BH model obtained using the Frobenius  method and  the agnostic parametrization: $\ell=2$, $M=1$. The columns "error" stand for the relative difference (in percents) between precise values of the QNMs and those obtained via the parametrization, while the "effect" columns designate the relative difference in percents between the values of QNMs for the Schwarzschild and quantum corrected black holes. }\label{tablecomp2}
\end{table*}
\section{Black hole shadows}

The fundamental equations for calculating the radius of a black hole shadow have been known for a long time \cite{Synge:1966okc,Claudel_2001} and have been applied in numerous studies (see, for example, \cite{Virbhadra:2002ju,Perlick:2021aok,Bisnovatyi-Kogan:2017kii,Perlick:2015vta,Konoplya:2019sns,Gan:2021xdl,Liu:2020ola} and references therein).

The radius of the circular photon orbit $r_{\rm ph}$ of static, spherically-symmetric spacetime can be defined as the solution of the following equation \cite{Synge:1966okc, Claudel_2001, Virbhadra:2002ju,Perlick:2021aok}
\begin{equation}
    \label{eq:photonGeod}
rf'(r)-2f(r)=0.
 \end{equation}
Then the shadow radius $R_{\rm sh}$ of the photon sphere $r_{\rm ph}$ observed by the distant observer is given by \cite{Synge:1966okc,PhysRevLett.125.141104}
\begin{equation}
\label{eq:shadow_radii}
R_{\rm sh}=\frac{r_{\rm ph}}{\sqrt{f(r_{\rm ph})}}.
\end{equation}
For the first model, we have
\begin{equation}
    (r-3 M) \left(4 M \xi ^2-r^3-2 \xi ^2 r\right)=0,
\end{equation}
which yields
\begin{equation}
\label{eq:rad_ph_sphere}
r_{\rm ph}=3M, \quad R_{\rm sh}=\frac{27 M^2}{\sqrt{\xi ^2+27 M^2}}.
\end{equation}
For the second model, we have  
\begin{equation}
r_{\rm ph}=3M,\quad R_{\rm sh}=3\sqrt{3}M,
\end{equation}
which coincides with the case of the Schwarzschild black hole. 

We can use the recent results of the  EHT observations of the black hole shadow to restrict the value of the quantum parameter $\xi$.

From observations of the shadow cast by Sgt $A^*$ black hole \cite{EventHorizonTelescope:2019dse,EventHorizonTelescope:2019ggy}, we have \cite{Vagnozzi_2023}
\eq{
4.55 M\lesssim R_{\rm sh}\lesssim   5.22M,\,\,(1\sigma).
}
This provides the following range for $\xi$ for the first model
\eq{
0\leq \xi\lesssim 2.866M.}
For the second model, $\xi$  cannot be restricted through the current black hole shadow observations
\eq{0\leq \xi< \infty.}

\section{Conclusions}\label{sec:conclusions}

While quasinormal modes of various models of quantum-corrected black holes have been extensively studied in recent years, no such studies have been conducted for the two recently proposed black hole models developed using the Hamiltonian constraints approach, which preserves general covariance \cite{Zhang:2024khj}.

In this work, we have demonstrated that the quasinormal modes for the two models of quantum-corrected black holes differ significantly from each other and from their classical (Schwarzschild) counterparts. However, there is a common feature for both models: while the fundamental mode changes slowly as the quantum parameter increases, the first few overtones deviate at an increasing rate from their Schwarzschild values, creating a characteristic "sound" of the event horizon. The real oscillation frequencies of the overtones rapidly approach zero as the quantum correction intensifies. The three independent methods used to find quasinormal modes—WKB approach, time-domain integration, and Frobenius method—are in very good agreement within the common range of their applicability. In the high-frequency (eikonal) regime, analytic formulas for quasinormal modes have been derived.
In addition to studying the branch of modes that are perturbative in the quantum parameter $\xi$, we observed frequencies that are non-perturbative in this parameter and exist at intermediate values of $\xi$. This non-perturbative branch of modes warrants more detailed study, especially if they are observed at sufficiently small values of $\xi$. We also found the quasinormal modes of axial gravitational perturbations of the black hole formed in the scenario of quantum-corrected gravitational collapse developed in~\cite{Lewandowski_2023}, and tested the agnostic parametrization of~\cite{Cardoso_2019} on all three models. It turns out that the accuracy of the parametrization depends on the model under consideration and is insufficient for overtones in all three models.

We have also calculated the radius of the shadow cast by both black hole models and demonstrated that, while the parameter $\xi$ in the first model could be constrained by observations of the Sgr A* black hole, the second model has a Schwazrschildian radius of shadows and does not allow for such a constraint.

Our work could be extended in several directions. While we focused on the most interesting bosonic perturbations, a similar analysis could be conducted for fermionic perturbations, particularly those describing neutrino perturbations. Additionally, the analytic formula derived in the eikonal limit could be extended to higher orders, providing a more accurate analytical approximation for the numerical data obtained in this study.

\begin{acknowledgments}
R. A. K.  acknowledges A. Zhidenko for useful discussions. O.S. is grateful to the Page family for their kind hospitality in Princeton.
\end{acknowledgments}

\newpage
\bibliographystyle{unsrt}
\bibliography{bibliography}
\end{document}